\documentclass[twocolumn]{aastex701}

\usepackage{siunitx}
\usepackage{comment}
\usepackage{CJKutf8} 
\usepackage{makecell}

\newcommand{\caltech}{Department of Astronomy, California Institute of Technology, Pasadena, CA 91125, USA}
\newcommand{\uclagps}{Department of Earth, Planetary, and Space Sciences, University of California, Los Angeles, CA 90095, USA}
\newcommand{\gps}{Division of Geological \& Planetary Sciences, California Institute of Technology, Pasadena, CA 91125, USA}
\newcommand{\ucsc}{Department of Astronomy \& Astrophysics, University of California, Santa Cruz, CA 95064, USA}
\newcommand{\keck}{W. M. Keck Observatory, 65-1120 Mamalahoa Hwy, Kamuela, HI, USA}
\newcommand{\ucla}{Department of Physics \& Astronomy, 430 Portola Plaza, University of California, Los Angeles, CA 90095, USA}
\newcommand{\jpl}{Jet Propulsion Laboratory, California Institute of Technology, 4800 Oak Grove Dr., Pasadena, CA 91109, USA}
\newcommand{\ucsd}{Department of Astronomy \& Astrophysics,  University of California, San Diego, La Jolla, CA 92093, USA}

\newcommand{\osu}{Department of Astronomy, The Ohio State University, 100 W 18th Ave, Columbus, OH 43210 USA}

\newcommand{\arizona}{James C. Wyant College of Optical Sciences, University of Arizona, Meinel Building 1630 E. University Blvd., Tucson, AZ 85721, USA}
\newcommand{\steward}{Steward Observatory, University of Arizona, 933 N Cherry Ave, Tucson, AZ, USA 85719}

\shorttitle{CD-35 2722~B KPIC}
\shortauthors{Wang et al.}

\begin{document}

\title{Chemical and Isotopic Homogeneity Between the L Dwarf CD-35 2722~B and its Early M Host Star}

\author[0000-0003-3092-4418]{Gavin Wang}
\email{gwang59@jhu.edu}
\affiliation{\caltech}
\affiliation{William H. Miller III Department of Physics \& Astronomy, Johns Hopkins University, Baltimore, MD 21218, USA}

\author[0000-0002-6618-1137]{Jerry W. Xuan}
\email{wxuan@caltech.edu}
\altaffiliation{51 Pegasi b Fellow}
\affiliation{\caltech}
\affiliation{\uclagps}

\author[0000-0001-9282-9462]{Dar\'io Gonz\'alez Picos}
\email{picos@strw.leidenuniv.nl}
\affiliation{Leiden Observatory, Leiden University, P.O. Box 9513, 2300 RA, Leiden, The Netherlands}

\author[0000-0002-3726-4881]{Zhoujian Zhang \begin{CJK*}{UTF8}{gbsn}(张周健)\end{CJK*}}
\email{zjzhang@rochester.edu}
\altaffiliation{NASA Sagan Fellow}
\affiliation{\ucsc}
\affiliation{Department of Physics \& Astronomy, University of Rochester, Rochester, NY 14627, USA}

\author[0000-0003-0097-4414]{Yapeng Zhang}
\email{yapzhang@caltech.edu}
\altaffiliation{51 Pegasi b Fellow}
\affiliation{\caltech}

\author[0000-0002-8895-4735]{Dimitri Mawet}
\email{dmawet@astro.caltech.edu}
\affiliation{\caltech}
\affiliation{\jpl}

\author[0000-0002-5370-7494]{Chih-Chun Hsu}
\email{chsu@northwestern.edu}
\affil{Center for Interdisciplinary Exploration and Research in Astrophysics (CIERA), Northwestern University, 1800 Sherman Ave, Evanston, IL 60201, USA}

\author[0000-0003-0774-6502]{Jason J. Wang \begin{CJK*}{UTF8}{gbsn}(王劲飞)\end{CJK*}}
\email{jason.wang@northwestern.edu}
\affil{Center for Interdisciplinary Exploration and Research in Astrophysics (CIERA), Northwestern University, 1800 Sherman Ave, Evanston, IL 60201, USA}
\affil{Department of Physics and Astronomy, Northwestern University, 2145 Sheridan Rd, Evanston, IL 60208, USA}

\author[0000-0003-0787-1610]{Geoffrey A. Blake}
\email{gab@gps.caltech.edu}
\affiliation{\gps}

\author[0000-0003-2233-4821]{Jean-Baptiste Ruffio}
\email{jruffio@ucsd.edu}
\affiliation{\ucsd}

\author[0000-0001-9708-8667]{Katelyn Horstman}
\email{khorstma@astro.caltech.edu}
\affiliation{\caltech}

\author[0000-0003-1399-3593]{Ben Sappey}
\email{bsappey@ucsd.edu}
\affiliation{\ucsd}

\author[0000-0002-6171-9081]{Yinzi Xin}
\email{yxin@caltech.edu}
\affiliation{\caltech}

\author[0000-0002-1392-0768]{Luke Finnerty}
\email{lfinnerty@astro.ucla.edu}
\affiliation{\ucla}

\author[0000-0002-1583-2040]{Daniel Echeverri}
\email{dechever@caltech.edu}
\affiliation{\caltech}

\author[0000-0001-5213-6207]{Nemanja Jovanovic}
\email{nem@caltech.edu}
\affiliation{\caltech}

\author[0000-0002-6525-7013]{Ashley Baker}
\email{abaker@caltech.edu}
\affiliation{\caltech}

\author{Randall Bartos}
\email{randall.d.bartos@jpl.nasa.gov}
\affiliation{\jpl}

\author[0000-0003-4737-5486]{Benjamin Calvin}
\email{bcalvin@astro.ucla.edu}
\affiliation{\caltech}
\affiliation{\ucla}

\author{Sylvain Cetre}
\email{scetre@keck.hawaii.edu}
\affiliation{\keck}

\author[0000-0001-8953-1008]{Jacques-Robert Delorme}
\email{jdelorme@keck.hawaii.edu}
\affiliation{\keck}

\author{Gregory W. Doppmann}
\email{gdoppmann@keck.hawaii.edu}
\affiliation{\keck}

\author[0000-0002-0176-8973]{Michael P. Fitzgerald}
\email{mpfitz@ucla.edu}
\affiliation{\ucla}

\author[0000-0002-4934-3042]{Joshua Liberman}
\email{jliberman@arizona.edu}
\affiliation{\arizona}
\affiliation{\steward}

\author[0000-0002-2019-4995]{Ronald A. L\'opez}
\email{rlopez@astro.ucla.edu}
\affiliation{\ucla}

\author[0000-0003-3165-0922]{Evan Morris}
\email{ecmorris@ucsc.edu}
\affiliation{\ucsc}

\author{Jacklyn Pezzato-Rovner}
\email{jpezzatorovner@gmail.com}
\affiliation{\caltech}

\author[0000-0001-5610-5328]{Caprice L. Phillips}
\email{cphillips@ucsc.edu}
\altaffiliation{NASA Sagan Fellow}
\affiliation{\ucsc}

\author{Tobias Schofield}
\email{toby.s117@gmail.com}
\affiliation{\caltech}

\author[0000-0001-6098-3924]{Andrew Skemer}
\email{askemer@ucsc.edu}
\affiliation{\ucsc}

\author[0000-0001-5299-6899]{J. Kent Wallace}
\email{james.k.wallace@jpl.nasa.gov}
\affiliation{\jpl}

\author[0000-0002-4361-8885]{Ji Wang \begin{CJK*}{UTF8}{gbsn}(王吉)\end{CJK*}}
\email{wang.12220@osu.edu}
\affiliation{\osu}

\begin{abstract}

CD-35 2722~B is an L dwarf companion to the nearby, $\sim 50-200$~Myr old M1 dwarf CD-35 2722~A. We present a detailed analysis of both objects using high-resolution ($R \sim 35,000$) $K$ band spectroscopy from the Keck Planet Imager and Characterizer (KPIC) combined with archival photometry. With a mass of $30^{+5}_{-4}~M_{\mathrm{Jup}}$ (planet-to-host mass ratio 0.05) and projected separation of $67\pm4$~AU from its host, CD-35 2722~B likely formed via gravitational instability. We explore whether the chemical composition of the system tells a similar story. Accounting for systematic uncertainties, we find $\mathrm{[M/H]}=-0.16^{+0.03}_{-0.02}\,\mathrm{(stat)} \pm 0.25\,\mathrm{(sys)}$~dex and $^{12}\mathrm{C}/^{13}\mathrm{C}=132^{+20}_{-14}$ for the host, and $\mathrm{[M/H]}=0.27^{+0.07}_{-0.06}\,(\mathrm{stat}) \pm 0.12\,(\mathrm{sys})$~dex, $^{12}\mathrm{CO}/^{13}\mathrm{CO}=159^{+33}_{-24}\,\mathrm{(stat)}^{+40}_{-33}\,\mathrm{(sys)}$, and $\mathrm{C/O} = 0.55 \pm 0.01\,(\mathrm{stat}) \pm 0.04\,(\mathrm{sys})$ for the companion. The chemical compositions for the brown dwarf and host star agree within the $1.5\sigma$ level, supporting a scenario where CD-35 2722~B formed via gravitational instability. We do not find evidence for clouds on CD-35 2722~B despite it being a photometrically red mid-L dwarf and thus expected to be quite cloudy. We retrieve a temperature structure which is more isothermal than models and investigate its impact on our measurements, finding that constraining the temperature structure to self-consistent models does not significantly impact our retrieved chemical properties. Our observations highlight the need for data from complementary wavelength ranges to verify the presence of aerosols in likely cloudy L dwarfs. 
\end{abstract}


\section{Introduction}

Brown dwarfs are substellar objects that occupy the mass range between the heaviest gas giant planets and the lightest stars, typically between 13 and 80 $M_\mathrm{Jup}$ \citep{Burrows_2001, Spiegel_2011, Dieterich_2014, Dupuy_2017, Schlaufman_2018}. These objects are unable to sustain stable hydrogen fusion in their cores, which results in their cooling over time and exhibiting a wide range of temperatures and atmospheric compositions. The study of brown dwarfs is thus crucial for understanding the boundary between planets and stars, as well as the processes that govern their formation and evolution. 

Among brown dwarfs, L dwarfs have temperatures ranging from approximately 1400 K to 2000 K \citep{Kirkpatrick_2005}, which is cool enough such that the atmospheres of L dwarfs are dominated by thick clouds, primarily thought to be composed of silicate and iron condensates \citep[see, e.g.,][]{Cushing_2006, Brock_2021, Suarez_2022, Miles_2023}. The presence of these clouds significantly influences their observed spectra and poses challenges for atmospheric modeling and characterization \citep{Burningham2017}. Among these challenges is the degeneracy between isothermal pressure-temperature profiles and clouds, and here we employ high-resolution spectroscopy and atmospheric retrievals to investigate this effect.

The chemical composition of brown dwarfs can give key insights into their formation \citep{Hoch_2023}. The C/O ratio and atmospheric metallicity, for instance, have long been used as indicators of formation pathway \citep{Oberg_2011}, with companions that are chemically similar to their host stars thought to form via gravitational instability \citep{Hoch_2023, xuan2024planets}. A recent notable addition is that of $^{13}$CO, an isotopologue that may also be indicative of formation location within the protoplanetary disk \citep{Zhang_2021}. In particular, objects which have lower $^{12}$CO/$^{13}$CO ratios than their host stars may have accreted additional $^{13}$CO from ice located beyond the CO snowline, which can be $^{13}$CO-enriched due to carbon fractionation processes including ion-exchange reactions and ice/gas isotopologue partitioning \citep{Langer_1984, Smith_2015, Zhang_2021, Lee_2024}. A caveat, however, is that chemical similarity between companions and host stars alone does not guarantee formation through gravitational stability \citep{Hsu2024}.

In this work, we perform a spectral analysis of CD-35 2722~B using high-resolution spectroscopy with the Keck Planet Imager and Characterizer (KPIC) in order to probe its bulk parameters, chemical composition, and cloud structure. This paper is organized as follows: \S \ref{sec:overview} provides an overview of the system, including the derivation of priors for the brown dwarf's physical properties based on evolutionary models. \S \ref{sec:data} describes the KPIC data and photometry, and \S \ref{sec:ccf} describes the cross-correlation functions obtained. In \S \ref{sec:analysis}, we describe our retrieval framework, including our forward model, pressure-temperature structure, and cloud models. \S \ref{hostresults} and \S \ref{compresults} present the results of the host star and companion. 
Finally, \S \ref{sec:discussion} discusses the implications of our results for the formation of CD-35 2722~B and we conclude in \S \ref{sec:conclusion}.

\section{System Overview} \label{sec:overview}

CD-35 2722~B is an L4.5 (NIR) dwarf that was discovered as part of the Gemini NICI Planet-Finding Campaign, which estimated its mass to be $31 \pm 8$ $M_{\mathrm{Jup}}$ \citep{Wahhaj_2011} based on its age, photometry, and evolutionary models. 
Its projected separation from the host star is $67 \pm 4$~AU, and it has an effective temperature of $\sim1900$~K \citep{Wahhaj_2011}. In all simulation scenarios considered by \citet{Boss_2023}, an object of this mass and separation cannot be formed by core accretion around an M dwarf, with massive disks of 0.05~$M_{\odot}$ only forming objects up to $\sim20~M_\mathrm{Jup}$. The goal of our work is to confirm whether the object's atmospheric composition reveals a consistent picture.

\subsection{Evolutionary Models} \label{subsec:evo}

\begin{figure*}[ht!]
    \centering
    \includegraphics[width=\linewidth]{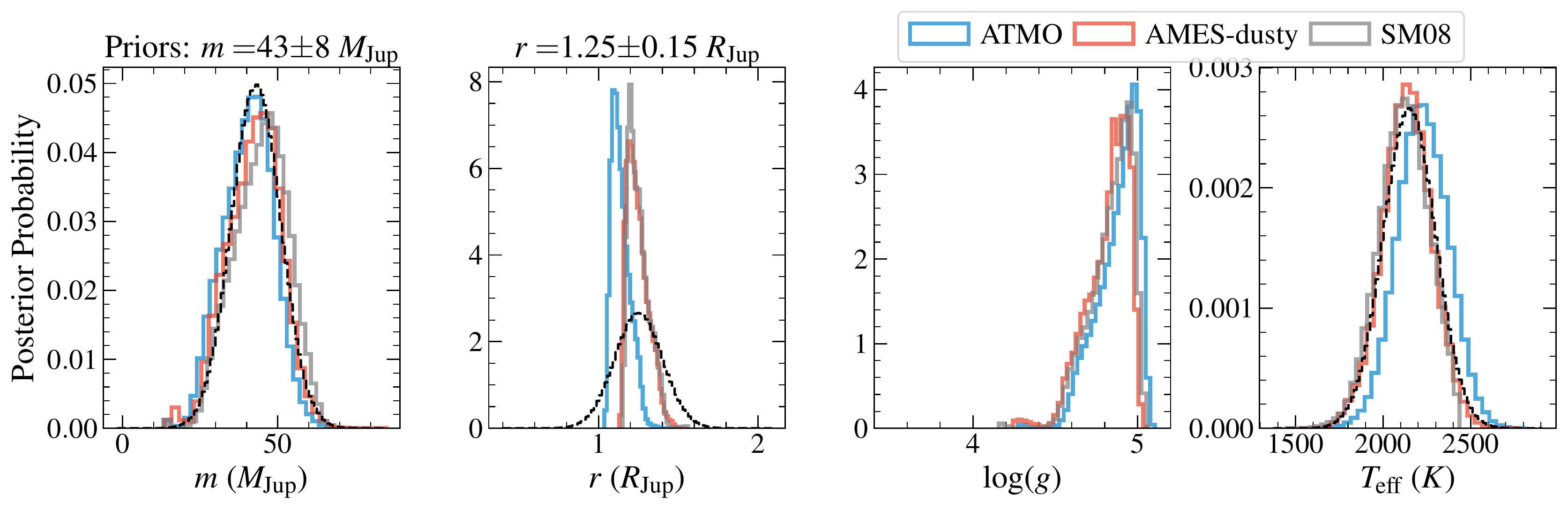}
    \caption{Results from evolutionary models, where the mass, radius, $\log g$, and $T_{\mathrm{eff}}$ are self-consistently calculated. We obtain the priors $43\pm 8~M_\mathrm{Jup}$ and $1.25\pm 0.15~R_\mathrm{Jup}$ (black dashed lines) on the mass and radius through visually comparing the Gaussians to the evolutionary model posterior probabilities. Note that the mass estimate is slightly larger than, but consistent with, the value of $31 \pm 8~M_\mathrm{Jup}$ from \citet{Wahhaj_2011}.}
    \label{fig:evo}
\end{figure*}

CD-35 2722~B does not have a well-constrained orbit or dynamical mass. Therefore, following existing works \citep[see, e.g., \S 2.2 of][]{xuan2024planets}, we derive priors on its bulk properties as expected from substellar evolutionary models. In particular, we estimate the mass, radius, surface gravity, and effective temperature of the brown dwarf given its age and luminosity. The host star is a part of the nearby AB Doradus association \citep{Torres_2008}, which is young, with estimated ages ranging from $50-200$~Myr \citep{Malo_2013, Gagne_2014, Bell_2015, Elliott_2016}. We adopt a uniform age prior of $50-200$~Myr and a luminosity of $\log (L_{\mathrm{bol}}/L_{\odot}) = -3.54 \pm 0.12$, obtained by correcting $\log (L_{\mathrm{bol}}/L_{\odot}) = -3.58 \pm 0.12$ from \citet{Wahhaj_2011} with the updated Gaia EDR3 parallax of $44.7203 \pm 0.0128$~mas \citep{Gaia_2021}. We use the ATMO \citep{Phillips_2020, Chabrier_2023}, AMES-Dusty \citep{Allard_2001}, and SM08 \citep{Saumon_2008} models; the results are shown in Figure \ref{fig:evo} and we adopt the priors $43\pm 8~M_\mathrm{Jup}$ and $1.25\pm 0.15~R_\mathrm{Jup}$ on the mass and radius. We refer the reader to \citet{xuan2024planets} for the details of this approach.

\section{Data} \label{sec:data}

\subsection{Keck Planet Imager and Characterizer}

KPIC is a high-contrast fiber-injection unit at the Keck II telescope \citep{Mawet_2016, Mawet_2017, Delorme2021, Jovanovic2025}, located downstream of the adaptive optics system and connected to NIRSPEC, a high-resolution ($R\sim35,000$) cross-dispersed echelle spectrograph \citep{McLean_1998, Martin_2018, Lopez_2020}. 

We observed the CD-35 2722 system on UT 2022 November 12 using KPIC science fiber 4. We obtained three 60-second exposures of the host star CD-35 2722~A and forty 120-second exposures of the brown dwarf CD-35 2722~B. We used a separation of $2813$~mas and position angle of 239.186$^{\circ}$ for the position of the brown dwarf relative to the host star. Additionally, we observed the A0V star HD 38056, which we used to derive the telluric model in our forward model (see \S \ref{subsec:fm}). We performed background subtraction, 1D spectral extraction, and wavelength calibration on the data.\footnote{\href{https://github.com/kpicteam/kpic_pipeline}{https://github.com/kpicteam/kpic\_pipeline}} We refer the reader to \citet{Wang_2021} for full details regarding the data reduction. 

$K$ band KPIC data contains a total of nine spectral orders from 1.95 to 2.49~$\mu$m. However, due to the strong CO$_2$ telluric lines in orders 37 to 39, and the wavelength calibration being imperfect for orders 34 to 36, we ended up using orders 31 to 33 (2.29 to 2.49~$\mu$m). These orders have the highest signal-to-noise ratio and cover the most important chemical features, including the CO bandhead in order 33 and CO and H$_2$O absorption features in all three orders. 

\subsection{Photometry}

Flux-calibrated photometric points in the MKO $J$, $H$, and $K$ bands were obtained by \citet{Wahhaj_2011} using the NICI instrument. These flux-calibrated values ($J = 13.63 \pm 0.11$, $H = 12.78 \pm 0.12$, and $K = 12.01 \pm 0.07$~mag) allow us to place constraints on the radius of the brown dwarf (which is not possible with only using high-pass filtered KPIC spectra), as well as offering a wider wavelength range that is more sensitive to clouds. We fit the photometry in conjunction with the KPIC data. 

\section{Cross-Correlation Functions} \label{sec:ccf}
\begin{figure}
    \centering
    \includegraphics[width=\linewidth]{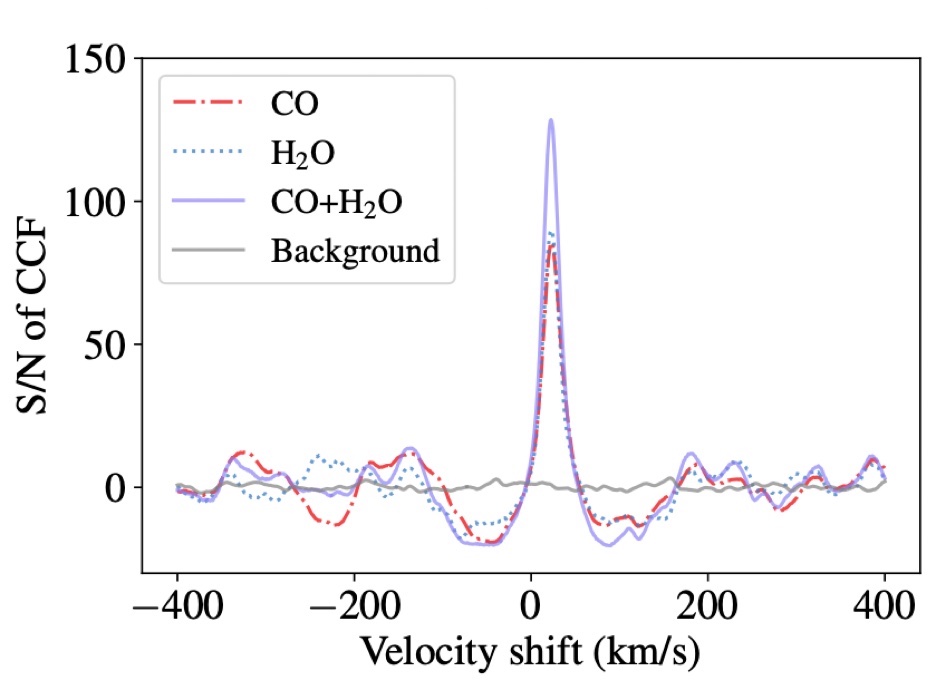}
    \caption{Cross-correlation function for the brown dwarf CD-35 2722~B. We find a highly significant $>100\sigma$ detection of this companion, which is expected given its brightness \citep[$K = 12.01 \pm 0.07$~mag;][]{Wahhaj_2011} and large separation (2.8'') from the host star. The gray line represents the CCF of background flux; its standard deviation is taken to be the CCF noise.}
    \label{fig:ccf}
\end{figure}

To determine the detection significance of CD-35 2722~B in our data, we cross-correlated the brown dwarf's spectrum with Sonora \citep{Marley_2021} model spectra containing CO and H$_2$O. We use models with $T_{\mathrm{eff}}=2000$~K and $\log g =4.5$. Results are shown in Figure \ref{fig:ccf}. The high cross-correlation function (CCF) signal-to-noise ($>100\sigma$) detection is expected due to the minimal spectral contamination by the host star some $3''$ away, and the brightness of the companion \citep[$K = 12.01 \pm 0.07$~mag from][]{Wahhaj_2011}. Additionally, given the high quality of our data, we searched for evidence for $^{13}$CO in the brown dwarf's spectrum. The results from this analysis are described in \S \ref{subsec:isotope}.

\section{Retrieval Framework} \label{sec:analysis}

Having gained a basic understanding of the data in hand, we next performed in-depth analyses using the most recent high-resolution ($R \sim 10^6$) line-by-line opacities in \texttt{petitRADTRANS} \citep{Molliere2019, Nasedkin2024}. Our retrieval framework performs nested sampling via \texttt{dynesty} \citep{Speagle_2020}. 

\subsection{Forward Modeling} \label{subsec:fm}

At each step of the retrieval, we downsample the $R \sim 10^6$ line opacities by a factor of 4 and convolve it to KPIC's resolution of $35,000$. The full forward modeling process, which takes into account the telluric and instrumental response and removes the continuum, is described in \citet{xuan2024planets}. 

\subsection{Chemistry}
We compute abundances by interpolating a pre-computed grid of chemical equilibrium abundances using [M/H] and C/O. We adopt the chemical equilibrium grid from \citet{Zhang2025}, who used easyCHEM \citep{2024arXiv241021364L} and extended the default petitRADTRANS grid \citep{Molliere2019} to higher temperatures while incorporating additional chemical species such as OH, Al, Ca, Fe, Mg, Si, and Ti. Our grid spans from 40 to 6000~K in temperature, $10^{-8}$ to $10^{3}$ bar in pressure, $-2$ to $3$~dex in [M/H], and 0.1 to 1.6 in C/O ratio. 

\subsection{Line Opacities} \label{subsec:opa}
We use high-temperature line opacities from \citet{xuan2024mdwarf} for $^{12}$CO and $^{13}$CO \citep{rothman2010}, H$_2$O \citep{polyansky2018}, TiO \citep{mckemmish2019}, VO \citep{mckemmish2016}, OH \citep{brooke2016}, and AlH \citep{yurchenko2018}, computed up to 4500~K, as well as H$_2$S \citep{azzam2016} computed to 2000~K. For atomic species, we include Na, K, Ca, Mg, Ti, Fe, Si, and Al from \citet{kurucz2011}. We also include H$_2$--H$_2$ \citep{Borysow_2001, Borysow_2002} and H$_2$--He \citep{Borysow_1988, Borysow_1989a, Borysow_1989b} collision-induced absorption opacities and $\mathrm{H}-$ bound-free and free-free continuum opacities \citep{Gray_2008}. 

\subsection{Pressure-Temperature Structure} \label{subsec:pt}
We use the same basic parameterization of the pressure-temperature ($P-T$) profile as described in \citet{xuan2024mdwarf} and adapted from \citet{piette2020}, which consists of an anchor temperature at a specified pressure and seven $\Delta T / \Delta P$ values between predefined pressure points, giving a total of eight parameters. The pressure points are chosen to capture the majority of the emission contribution, and are set at $\log_{10} P~(\mathrm{bar}) =$ 1.0, 0.5, 0.2, $-0.1$, $-0.4$, $-1.0$, $-2.0$, and $-4.7$. Under this parameterization, we implement two different treatments of the $P-T$ profile, as described below.

In the first case, we apply Gaussian priors on these eight $P-T$ parameters. Our Gaussian priors are derived from the cloudy, self-consistent Sonora Diamondback models \citep{morley2024sonora}, using a procedure similar to that in \citet{Zhang_2023}. In particular, we choose a set of models close to the expected properties of the brown dwarf ($T_{\mathrm{eff}}=1700-1900$~K, $\log g=4.5-5.5$) and calculate the associated values of each $P-T$ parameter. We define $\sigma$ for each parameter such that all values are encompassed within $\pm 2 \sigma$ of the median. A difference from \citet{Zhang_2023} is that whereas they place priors on $d\ln T/ d\ln P$, we place priors on $\Delta T / \Delta P$. The Diamondback model was selected because it is the latest grid model with a self-consistent treatment of clouds, which are expected to be important for CD-35 2722~B. 

\begin{table*}[ht!]
    \centering
    \begin{tabular}{lccccc}
    \hline \hline 
         $P-T$ Treatment & Cloud Model & Particle Type & Species & $\ln \mathcal{Z}$ & $\Delta \ln \mathcal{Z}$ \\
         \hline
         \textbf{Gaussian priors} & \textbf{Clear} & \textbf{--} & \textbf{--} & $\mathbf{-13804.896}$ & \textbf{0.0} \\
         Gaussian priors & Gray opacity & -- & -- & $-13804.777$ & $+0.119$ \\
         Gaussian priors & EddySed & Amorphous & MgSiO$_3$$+$Fe & $-13805.139$ & $-0.243$ \\
         Gaussian priors & EddySed & Amorphous & MgSiO$_3$ & $-13802.673$ & $+2.223$ \\
         Gaussian priors & EddySed & Amorphous & Fe & $-13804.025$ & $+0.871$ \\
         Gaussian priors & EddySed & Crystalline & MgSiO$_3$$+$Fe & $-13804.663$ & $+0.233$ \\
         Gaussian priors & EddySed & Crystalline & MgSiO$_3$ & $-13805.060$ & $-0.164$ \\
         Gaussian priors & EddySed & Crystalline & Fe & $-13805.341$ & $-0.445$ \\
         Fixed & Clear & -- & -- & $-13904.643$ & $-99.747$ \\
         Fixed & Gray opacity & -- & -- & $-13896.277$ & $-91.381$ \\
         Fixed & EddySed & Amorphous & MgSiO$_3$ & $-13859.334$ & $-54.438$ \\
         \hline
    \end{tabular}
    \caption{Log-evidences of retrievals, with the $P-T$ treatment labeled as either using Gaussian priors or fixing to the cloudy Diamondback models. All retrievals use equilibrium chemistry. We take the cloud-free model as the baseline to determine the Bayes factor of each model; it is also the model we use in our final fit.}
    \label{tab:lnz}
\end{table*}

In our second, alternative $P-T$ treatment, we fix the $P-T$ profile to that in the Sonora Diamondback grid \citep{morley2024sonora} which was closest to the best-fit profile from our previous model. In other words, here we do not fit for the $P-T$ at all. The purpose of this was to gauge how sensitive our derived chemical parameters were to variations in the $P-T$ structure, as well as to explore a potential method of mitigating isothermality. We fixed the $P-T$ profile to the model corresponding to $T_{\mathrm{eff}}=1500$~K, $\log g=4.5$, sedimentation efficiency $f_{sed}=1$, and $\mathrm{[M/H]}=0$, while slightly extrapolating the profile onto the larger pressure range of $10^{-4.7}$ to $10^{1.0}$ bars used in our retrievals. We compare the Bayesian evidences of the two $P-T$ treatments (fitting vs fixing) in Table \ref{tab:lnz}. 

\subsection{Cloud Models}

Following \citet{Xuan_2022}, we use the EddySed cloud model from \citet{Ackerman_2001}. This model allows the specification of the cloud particle type, which are either irregularly shaped (crystalline) particles and modeled as a distribution of hollow spheres, or spherical and homogeneous (amorphous) particles with cross-sections modeled with Mie theory \citep{Min_2005, Molliere2019}. 

For each cloud species, we fit for the mass fraction of the cloud at its base (at the intersection with the retrieved $P-T$ profile), expressed in terms of $\log (f_{\mathrm{retrieved}} / f_\mathrm{models})$, where $f_\mathrm{models}$ is the cloud mass fraction predicted by the easyCHEM model \citep{2024arXiv241021364L} given a set of C/O and [M/H]. We also fit for the sedimentation efficiency $f_\mathrm{sed}$, separately for each cloud species, which partially determines the extent of the clouds, with smaller values indicating more vertically extensive clouds. We fit for the log-scale sedimentation efficiency, as our preliminary analyses indicated low values. Finally, we fit for the vertical eddy diffusion constant $\log_{10} (K_{zz})$, the carbon quench pressure $\log_{10} (P_{\mathrm{quench}})$, and the $\sigma$ of the lognormal cloud particle size distribution $\sigma_{\mathrm{norm}}$. 

Condensate species expected in brown dwarf atmospheres include MgSiO$_3$, Fe, Na$_2$S, and KCl \citep{Marley_2013}, but only the first two have intersections with the $P-T$ structure of CD-35 2722~B. We thus place our focus on these species, as the others are not expected to form clouds in the atmosphere of the companion. 

\subsection{Host Star}

Due to the host star's high effective temperature of $\sim 3700$~K \citep{Gaia_2021}, we model its spectrum via a custom grid of PHOENIX models that account for varying carbon isotope ratios \citep{Husser2013, Zhang_2024, GonzalezPicos2024}. The fundamental stellar parameters---effective temperature, surface gravity, and metallicity---are determined through fitting. Our model grid spans effective temperatures ranging from 3500 to 3900~K, $\log g$ from 3.5 to 5.5, metallicity from $-0.5$ to 0.5~dex, and carbon isotope ratios $^{12}\mathrm{C}/^{13}\mathrm{C}$ ranging from 31 to 196. We use the solar reference from \citet{Asplund_2009}. To forward-model spectra from this four-dimensional grid, we use the \texttt{Starfish} framework, which facilitates probabilistic spectral interpolation \citep{Czekala_2015}. Additionally, we fit the star’s radial velocity and incorporate rotational broadening by convolving the model spectra with a rotational kernel\footnote{\href{https://github.com/sczesla/PyAstronomy}{https://github.com/sczesla/PyAstronomy}} \citep{Czesla_2019}. The projected rotational velocity, $v \sin i$, is treated as a free parameter with a uniform prior from 1 to 30~km/s, and we allow the radial velocity to vary from $-20$ to $+40$~km/s. We fix the linear limb darkening coefficient to 0.19, the theoretically calculated value from the Limb Darkening Toolkit\footnote{\href{https://github.com/hpparvi/ldtk}{https://github.com/hpparvi/ldtk}} \citep{Parviainen2015} for a star of $T_{\mathrm{eff}} = 3700$~K, $\log g = 4.3$, and $\mathrm{[M/H]}=-0.1$~dex. The instrumental line spread function is accounted for by convolving the model spectra with a Gaussian kernel based on the measured line widths of the fiber, and we remove data due to strong tellurics using a transmission threshold of 0.30, along with the 20 pixels around them. This masks 18\% of the data points; using a smaller threshold of 0.10 masks 12\% of data and yields worse but consistent results for all parameters, so we choose a threshold of 0.30 for robustness. We forward-model the continuum of the model using a spline decomposition as defined in \citet{Ruffio_2023} and implemented in \citet{GonzalezPicos2024}. 

For parameter inference, we perform Bayesian retrievals using the \texttt{PyMultinest} \citep{Buchner2014} implementation of the \texttt{Multinest} algorithm \citep{Feroz_2009} with importance nested sampling \citep{Feroz2019}. Each retrieval is run with 400 live points, a constant efficiency of 5\%, and an evidence tolerance of 0.5. At each likelihood evaluation, we determine the optimal linear parameters for the spline model and rescale the uncertainty following \citet{Ruffio_2019}.

\begin{deluxetable}{@{\extracolsep{0pt}}lc@{}}[h!]
\tablecaption{Posteriors for the CD-35 2722 A grid fit. The uncertainties reported here include both statistical and systematic uncertainties for $T_{\mathrm{eff}}$, $\log g$, and $\mathrm{[M/H]}$, which are separately reported and labeled as such.} \label{tab:star}
\tablehead{
\colhead{Parameter} & \colhead{Posterior}
}
\startdata
$T_{\mathrm{eff}}$ (K) & $3689^{+4}_{-3}\,\mathrm{(stat)} \pm 100\,\mathrm{(sys)}$  \\
$\log g$        & $4.24 \pm 0.03\,\mathrm{(stat)} \pm 0.50\,\mathrm{(sys)}$   \\
$\mathrm{[M/H]}$ (dex)            & $-0.16 ^{+0.03}_{-0.02}\,\mathrm{(stat)} \pm 0.25\,\mathrm{(sys)}$ \\
$^{12}$C/$^{13}$C  & $132^{+20}_{-14}$     \\
Radial Velocity (km/s)        & $32.28\pm0.03$  \\
$v \sin i$ (km/s)   & $11.95 \pm 0.05$  \\
Limb darkening coefficient & Fixed, 0.19 \\
\enddata
\end{deluxetable}

\section{Host Star Results} \label{hostresults}

\begin{figure*}
    \centering
    \includegraphics[width=0.95\linewidth]{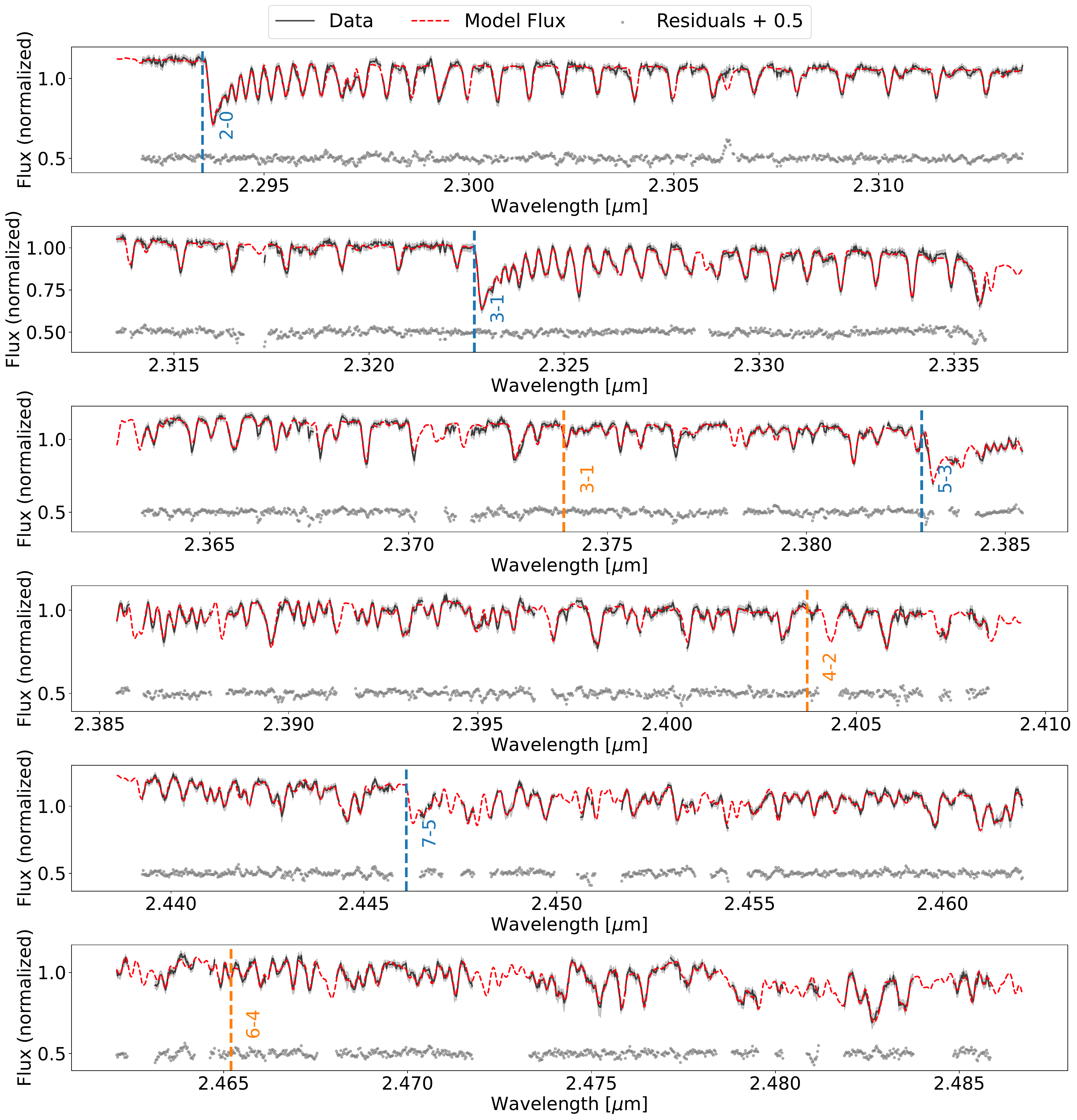}
    \caption{Best-fit model (red) and spectrum (black, with uncertainties as gray shaded regions) for CD-35 2722~A, using the PHOENIX grid. Residuals have been shifted by $+0.5$ for visual purposes. The corresponding properties are shown in Figure \ref{fig:host_corner}. The gaps in the data are due to strong tellurics, which have been masked based on the response function with a transmission threshold of 0.30. The bandheads of $^{12}$CO and $^{13}$CO are shown as vertical dashed lines, in blue and orange respectively. The labels denote the transitions between vibrational states. The spectral feature at 2.307~$\mu$m is due to a linelist mismatch or due to an unidentified species not present in the host star; this highlights a limitation of precomputed grid models. The standard deviation of the residuals (0.018) is comparable to the median errorbar (0.021).}
    \label{fig:host_spec}
\end{figure*}

\begin{figure*}
    \centering
    \includegraphics[width=0.9\linewidth]{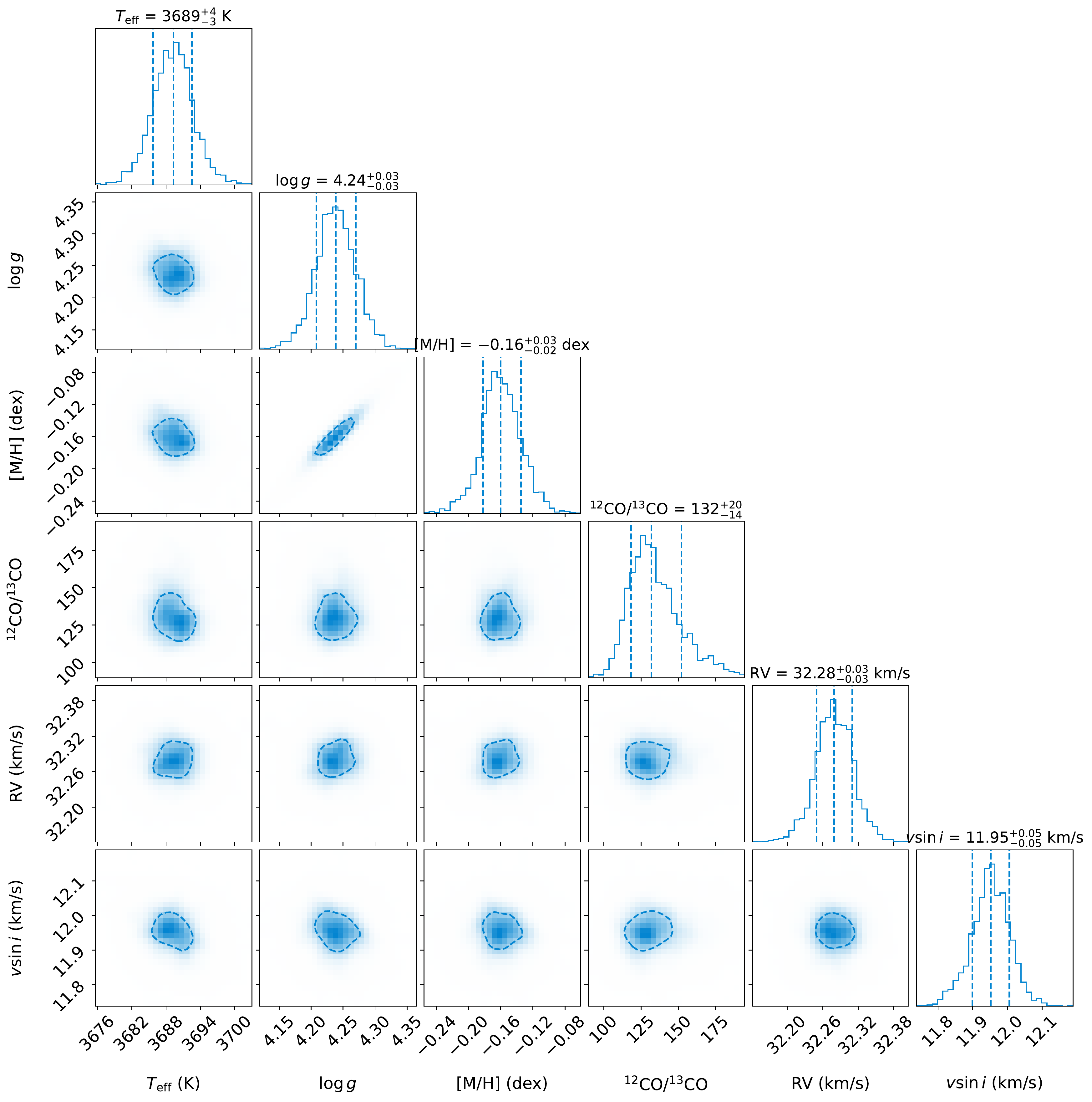}
    \caption{Joint posterior distribution for best-fit parameters to CD-35 2722~A spectrum. Note that the uncertainties here do not include systematic uncertainties, which inflate the errors for $T_{\mathrm{eff}}$ and $\mathrm{[M/H]}$.}
    \label{fig:host_corner}
\end{figure*}

The best-fit spectra and corner plot for the host star are shown in Figures \ref{fig:host_spec} and \ref{fig:host_corner}. We find a subsolar metallicity of $-0.16^{+0.03}_{-0.02}$~dex and a $^{12}$C/$^{13}$C ratio of $132^{+20}_{-14}$. The effective temperature of the M dwarf host is $3689^{+4}_{-3}$~K. We note that the quoted uncertainties only include statistical and not systematic uncertainties. There could be systematic uncertainties from the models, for example due to the lack of flexibility in the pressure structures of the Phoenix grid. The spacing of the grid is $100$~K in temperature, 0.5 in $\log g$, and 0.25~dex in metallicity \citep{GonzalezPicos2024}, which we conservatively estimate as our systematic errors. The full posteriors are listed in Table \ref{tab:star}. For the purposes of comparing host and companion, we add the uncertainties in quadrature to obtain a metallicity of $-0.16 \pm 0.25$~dex and an effective temperature of $3689 \pm 100$~K. 

\section{Companion Results} \label{compresults}
\subsection{Temperature Structure and Adopted Model}

\begin{figure}[ht!]
    \centering
    \includegraphics[width=\linewidth]{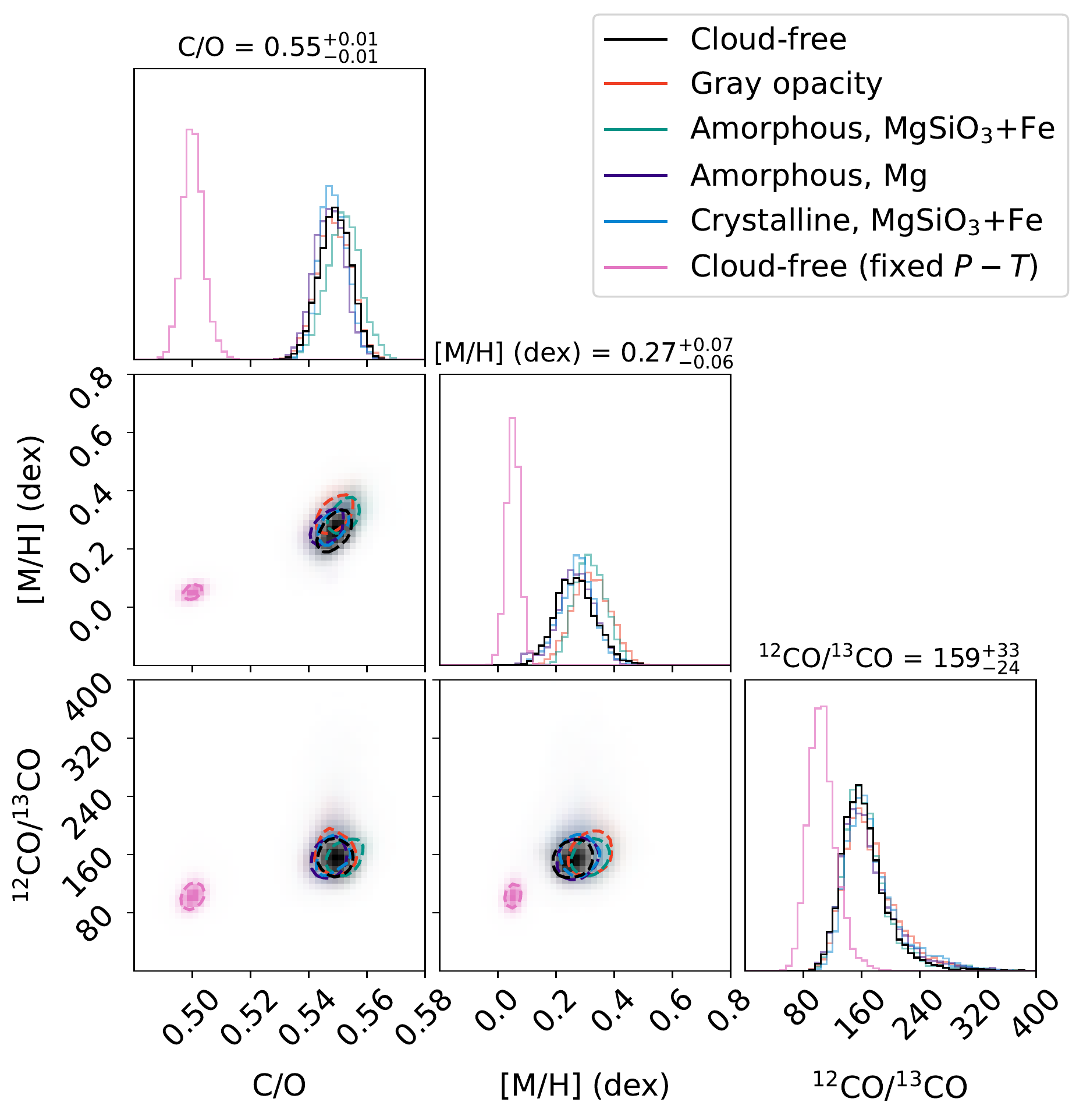}
    \caption{Posterior distribution of C/O, $\mathrm{[M/H]}$, and $^{12}$CO/$^{13}$CO for select retrievals in Table \ref{tab:lnz}. All retrievals use equilibrium chemistry and Gaussian priors on the $P-T$ profile, except for the one shown in pink which fixes the $P-T$ profile and is labeled as such. The uncertainties here are only statistical and do not include systematic uncertainties. All parameters are consistent to within $3\sigma$ across the models except for C/O, which is nominally discrepant at $7\sigma$ between the fixed $P-T$ model and all other models. This discrepancy lowers to $0.9\sigma$ upon considering systematic uncertainties. The cloud-free model with Gaussian $P-T$ priors, represented in black, is the configuration we choose as our fiducial result.}
    \label{fig:corners}
\end{figure}

As described in \S \ref{subsec:pt}, we performed retrievals using free and fixed pressure-temperature structures, as well as a combination of cloud models. These models are listed in Table \ref{tab:lnz} and key chemical indicators for select models are shown in Figure \ref{fig:corners}. 

\begin{figure*}[hbt!]
    \centering
    \includegraphics[width=0.9\linewidth]{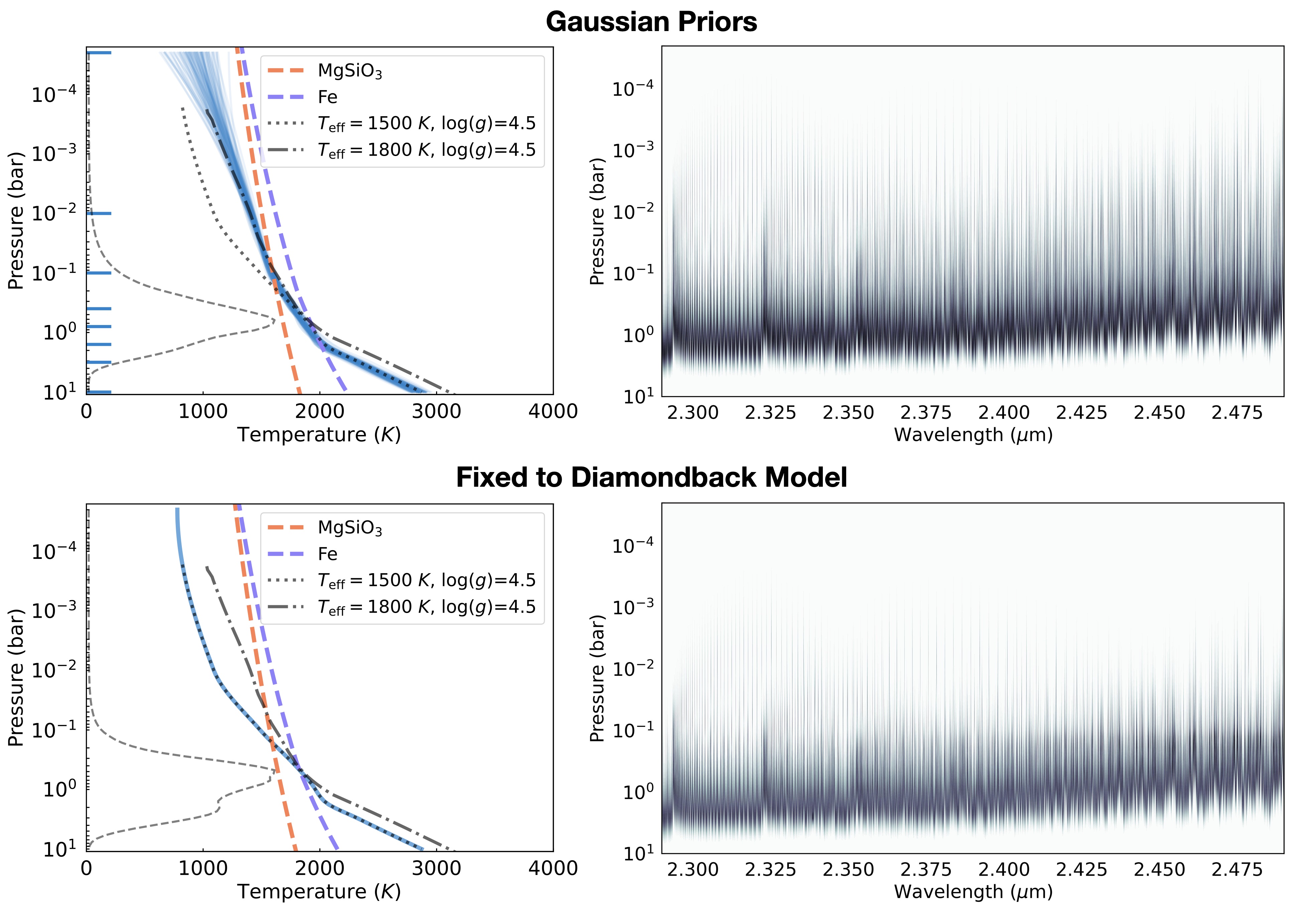}
    \caption{\textbf{Left:} Retrieved pressure-temperature profile (blue) with cloud condensate curves overplotted for enstatite (MgSiO$_3$) and iron (Fe). The overplotted Diamondback models (dotted and dash-dotted lines) each have $f_{\mathrm{sed}}=1$. The top row corresponds to the fit with Gaussian priors on the $P-T$ profile (pressure points shown as blue ticks), and the bottom row shows the fit fixing the $P-T$ profile to the $T_{\mathrm{eff}}=1500$~K Sonora Diamondback model. We adopt the former fit as the baseline model in this work, although it retrieves temperature structures that are slightly isothermal compared to the $T_{\mathrm{eff}}=1500$~K Diamondback model and hotter by about $300$~K in the upper parts of the atmosphere. The upper atmosphere instead matches the $T_{\mathrm{eff}=1800}$~K Diamondback model, demonstrating that a single model cannot match both the lower and upper atmosphere. \textbf{Right:} Emission contribution functions for the corresponding models, where darker colors represent a higher fraction of the emergent flux originating from the corresponding pressure level. Gray dashed lines are the contribution functions integrated over all wavelengths, showing that much of the contribution comes from above the Fe cloud deck pressure, especially in the Gaussian prior case.}
    \label{fig:ptp}
\end{figure*}

\begin{figure*}
    \centering
    \includegraphics[width=\linewidth]{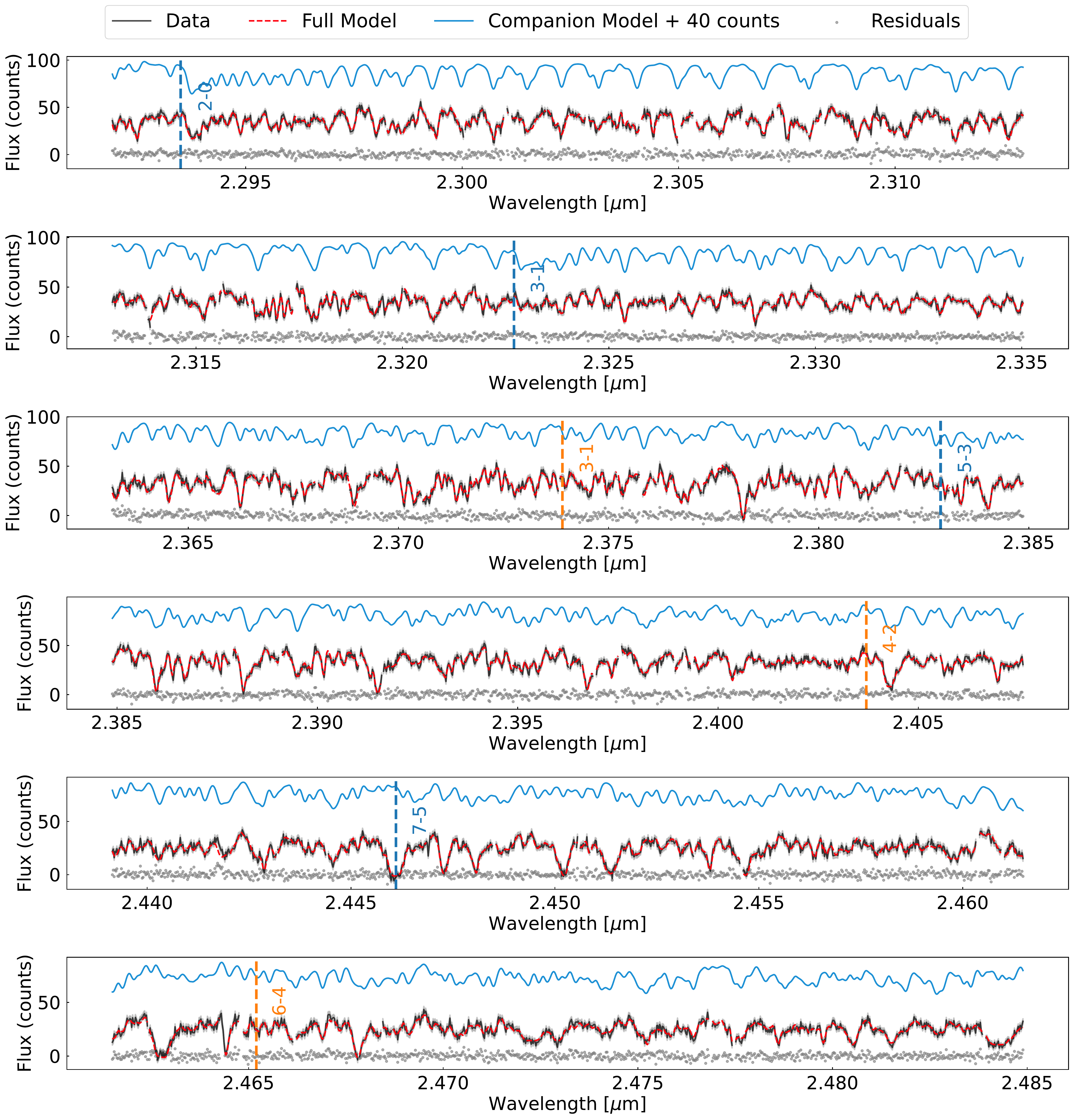}
    \caption{KPIC spectrum and model calculated according to the best-fit retrieved parameters. The data are shown as a black line with $1\sigma$ uncertainties enclosed in gray, the full forward model is shown as the red line, and the companion model before forward modeling is shown as a blue line. The residuals are represented by gray points. The bandheads of $^{12}$CO and $^{13}$CO are shown as vertical dashed lines, in blue and orange respectively. The labels denote transitions between vibrational states. }
    \label{fig:kpic}
\end{figure*}

\begin{figure}
    \centering
    \includegraphics[width=\linewidth]{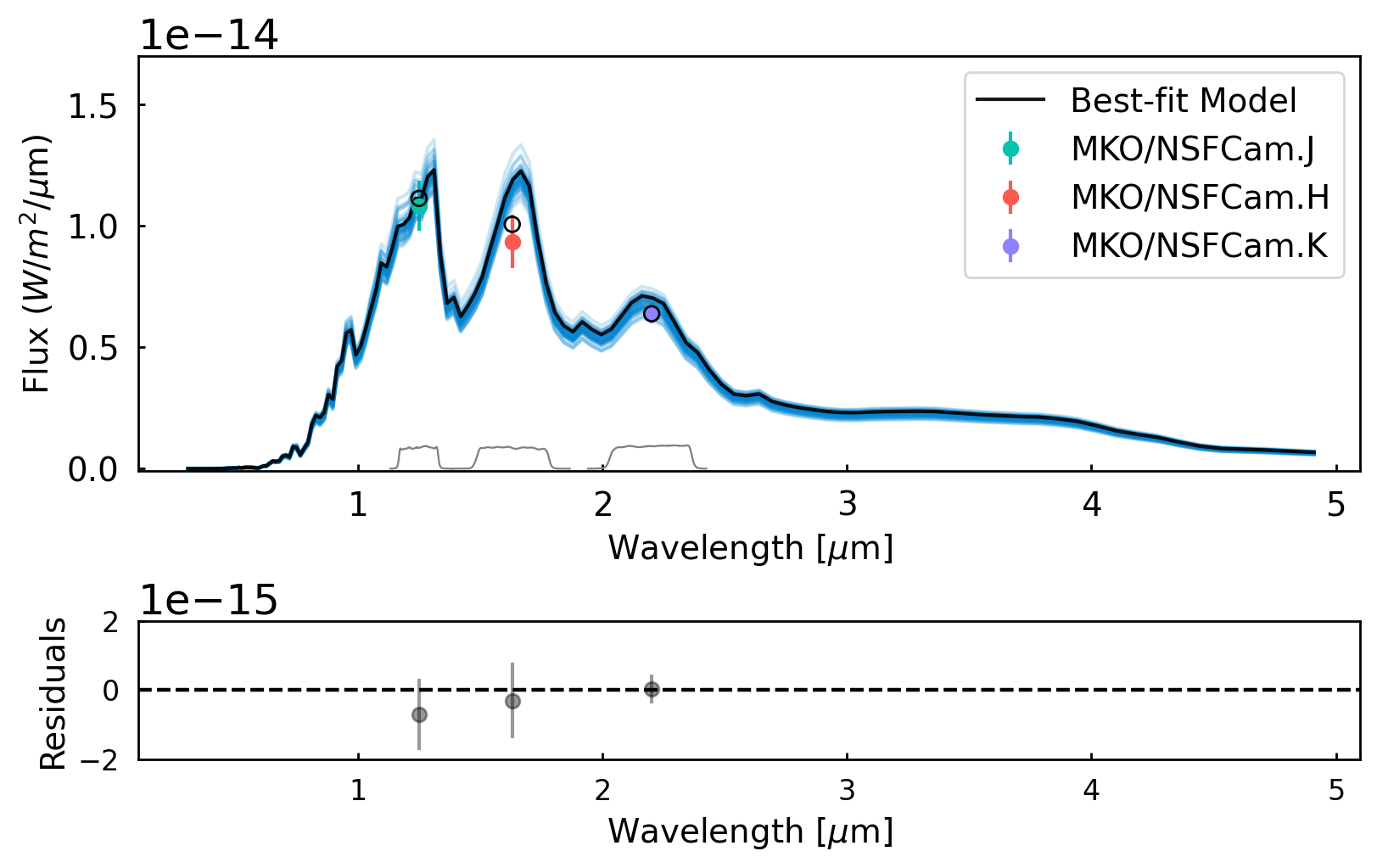}
    \caption{The same best-fit model as represented in Figure \ref{fig:kpic}, but for the $J$, $H$, and $K$ photometry. The colored points represent the measured values and the black points and curve represent the model. The blue lines represent 100 random draws of the best-fit models to the brown dwarf's spectrum. Filter transmission curves are shown in gray.}
    \label{fig:phot}
\end{figure}

We adopt the clear model with Gaussian priors on the $P-T$ profile as our final model because it has the highest log-evidence, although it is slightly isothermal. We refer the reader to \S \ref{subsec:iso} for an in-depth discussion of this decision. The posteriors from this retrieval are shown in Table \ref{tab:comp}, along with those from the clear retrieval with a fixed $P-T$ profile for comparison. These are labeled  ``Gaussian'' and ``Fixed'' respectively.  The best-fit $P-T$ profiles and emission contribution functions for each $P-T$ treatment is shown in Figure \ref{fig:ptp}. Finally, the forward-modeled spectra and photometry corresponding to our adopted model are shown in Figures \ref{fig:kpic} and \ref{fig:phot}. 

\begin{deluxetable*}{@{\extracolsep{0pt}}lccc@{}}
\tablecaption{Prior and posterior values for retrieval (1) with Gaussian priors on the $P-T$ structure ($\ln \mathcal{Z} = -13804.896$), and (2) with fixing the $P-T$ to the closest-matching model within the Sonora Diamondback grid ($\ln \mathcal{Z} = -13904.643$). We adopt the fit with Gaussian priors as final because of its higher log-evidence, but include the latter to demonstrate how much our retrieved results are affected by changes in the $P-T$ structure. The statistical and systematic uncertainties are reported separately for $\mathrm{C/O}$, $\mathrm{[M/H]}$, and $^{12}$CO/$^{13}$CO. Additionally, an oxygen correction has been applied to the $\mathrm{C/O}$ ratios following \S \ref{subsec:oxygen}.} \label{tab:comp}
\tablehead{
\colhead{Parameter Name} & \colhead{Priors} & \colhead{\textbf{Gaussian}} & \colhead{Fixed}
}
\startdata
\textbf{Bulk Parameters}\\
Mass ($M_\mathrm{Jup}$)         &   $\mathcal{N}^{[1]} (43, 8)$     & $29.5^{+5.1}_{-4.3}$ & $37.7 \pm 2.5$ \\
Radius ($R_\mathrm{Jup}$) &    $\mathcal{N} (1.25, 0.15)$    & $1.55^{+0.05}_{-0.06}$ & $1.35 \pm 0.04$ \\
Radial Velocity (km/s)      &  $\mathcal{U}^{[2]} (-70, 130)$      & $32.91 \pm 0.08$  & $32.84^{+0.07}_{-0.06}$  \\
$v \sin i$ (km/s)           &   $\mathcal{U} (0, 100)$     & $9.58^{+0.25}_{-0.24}$  & $10.61^{+0.18}_{-0.17}$  \\
\hline \textbf{Instrumental Parameters}\\
Order 37 flux (counts)       &   $\mathcal{U} (0, 150)$     & $53.70^{+0.51}_{-0.50}$ & $52.81 \pm 0.54$ \\
Order 38 flux (counts)      &   $\mathcal{U} (0, 150)$     & $49.75^{+0.35}_{-0.31}$ & $49.20 \pm 0.37$ \\
Order 39 flux (counts)      &   $\mathcal{U} (0, 150)$     & $41.69^{+0.33}_{-0.30}$  & $40.90 \pm 0.32$  \\
Flux error inflation        &    $\mathcal{U} (1, 4)$    & $1.81 \pm 0.02$  & $1.84 \pm 0.02 $  \\
Line spread function scale  &   $\mathcal{U} (1.0, 1.2)$     & $1.16^{+0.03}_{-0.05}$  & $1.18^{+0.01}_{-0.03}$  \\
\hline \textbf{$\mathbf{P-T}$ Parameters}\\
$T_\text{anchor}$ [$\log_{10}(P) = -0.1$] (K)       &  $\mathcal{N}(1920, 60)^*$      & $1886^{+23}_{-22}$ & -- \\
$\Delta T_1$ [1.0 to 0.5] (K)            &  $\mathcal{N}(530, 15)^*$      & $528^{+12}_{-14}$ & -- \\
$\Delta T_2$ [0.5 to 0.2] (K)            &  $\mathcal{N}(310, 20)^*$      & $301^{+17}_{-19}$ & -- \\
$\Delta T_3$ [0.2 to $-0.1$] (K)            &  $\mathcal{N}(180, 75)^*$      & $137^{+27}_{-30}$ & -- \\
$\Delta T_4$ [$-0.1$ to $-0.4$] (K)            &  $\mathcal{N}(125, 40)^*$      & $113^{+21}_{-20}$  & -- \\
$\Delta T_5$ [$-0.4$ to $-1.0$] (K)            &  $\mathcal{N}(200, 40)^*$      & $183^{+23}_{-21}$  & --  \\
$\Delta T_6$ [$-1.0$ to $-2.0$] (K)            &  $\mathcal{N}(200, 50)^*$      & $176^{+30}_{-29}$  & -- \\
$\Delta T_7$ [$-2.0$ to $-4.7$] (K)            &  $\mathcal{N}(400, 170)^*$      & $493^{+118}_{-108}$  & -- \\
\hline \textbf{Chemical Parameters}\\
C/O                         &  $\mathcal{U}(0.1, 1.0)$      & $0.55 \pm 0.01\,(\mathrm{stat}) \pm 0.04\,(\mathrm{sys})^{[3]}$  & $0.50 \pm 0.01\,(\mathrm{stat})\pm 0.04\,(\mathrm{sys})^{[4]}$ \\ 
$\mathrm{[M/H]}$ (dex)     &    $\mathcal{U}(-1.5, 1.5)$    & $0.27^{+0.07}_{-0.06}\,(\mathrm{stat}) \pm 0.12\,(\mathrm{sys})$   & $0.05 \pm 0.02\,(\mathrm{stat}) \pm 0.12\,(\mathrm{sys})$   \\
$\log_{10}(^{12}\mathrm{CO}/^{13}\mathrm{CO})$         &  $\mathcal{U}(0, 6)$      & $2.20^{+0.08}_{-0.07}\,(\mathrm{stat}) \pm 0.10\,(\mathrm{sys})$   & $2.02 \pm 0.08\,(\mathrm{stat})\pm 0.10\,(\mathrm{sys})$   \\
$\log_{10}(P_\mathrm{quench})$    &   $\mathcal{U}(-4.7, 1.0)$     & $0.33^{+0.44}_{-0.62}$  & $-3.01^{+1.22}_{-1.15}$  \\
\enddata
\begin{itemize}
        \item[] $^{[1]}$ $\mathcal{N} (a, b)$ represents a Normal prior with median $a$ and standard deviation $b$
        \item[] $^{[2]}$ $\mathcal{U} (a, b)$ represents a uniform prior between $a$ and $b$
        \item[] $^{[3]}$ Before oxygen correction: $0.69 \pm 0.01\,(\mathrm{stat})\pm 0.04\,(\mathrm{sys})$
        \item[] $^{[4]}$ Before oxygen correction: $0.61 \pm 0.01\,(\mathrm{stat})\pm 0.04\,(\mathrm{sys})$
        \item[] $^*$ Not applicable to retrieval fixing the pressure-temperature profile to a Sonora model
    \end{itemize}
\end{deluxetable*}

\subsection{No Conclusive Detection of Clouds} \label{subsec:noclouds}

We find no significant evidence for MgSiO$_3$ or Fe clouds in CD-35 2722~B using Gaussian priors on our $P-T$ profile (see Table \ref{tab:lnz}), despite the object being expected to have clouds as a photometrically red mid-L dwarf, as well as having a $P-T$ profile that intersects with the condensation curves for both cloud species. Among the models which fix the $P-T$ structure to a Sonora Diamondback model, we find significant preference for those which include clouds, with gray opacity and EddySed clouds preferred over no clouds by $\Delta \ln\mathcal{Z}=8.4$ and $45.3$ respectively. However, all of these models have much lower log-evidence ($\Delta \ln\mathcal{Z}<-50$) than our adopted model, which is the clear model with Gaussian $P-T$ priors. We thus rule out a definitive detection of clouds. 

\subsection{Retrieved Chemical Properties} \label{subsec:oxygen}

Since our retrievals use equilibrium chemistry, we use C/O and $\mathrm{[M/H]}$ to parametrize the chemical abundances required for computing opacities. We retrieve C/O values of $0.69 \pm 0.01$ when using Gaussian priors on the pressure-temperature structure and $0.61 \pm 0.01$ for the fixed $P-T$ profile. However, it is likely that some of the oxygen in the brown dwarf has been trapped in its MgSiO$_3$ clouds \citep{Xuan_2022, Nasedkin_2024, Calamari_2024}, even though we don't conclusively detect such clouds in our data. We thus performed an oxygen correction to all of our C/O values using Equation 12 in \citet{Calamari_2024}. 

Applying the correction yields a nearly solar C/O ratio of $0.55 \pm 0.01$ for the brown dwarf companion. Additionally folding in a systematic uncertainty of 0.04 in C/O, which was estimated from KPIC datasets on a similarly high S/N target \citep{xuan2024mdwarf}, we obtain $\mathrm{C/O}=0.55\pm 0.01\,(\mathrm{stat})\pm0.04\,(\mathrm{sys})$, which we report as our final measurement. All C/O values reported elsewhere in this paper are after applying the oxygen correction.

We also measure $\mathrm{[M/H]}=0.27^{+0.07}_{-0.06}$~dex and $^{12}\mathrm{CO}/^{13}\mathrm{CO}=159^{+33}_{-24}$ for the companion, which when accounting for systematic uncertainties of 0.12~dex in $\mathrm{[M/H]}$ and 0.10~dex in $^{12}$CO/$^{13}$CO \citep{xuan2024mdwarf} yields $\mathrm{[M/H]}=0.27^{+0.07}_{-0.06}\,\mathrm{(stat)} \pm 0.12\,\mathrm{(sys)}$~dex and $^{12}\mathrm{CO}/^{13}\mathrm{CO}=159^{+33}_{-24}\,\mathrm{(stat)}^{+40}_{-33}\,\mathrm{(sys)}$. These are listed in Table \ref{tab:comp}. For the purposes of our following analyses, we add the statistical and systematic uncertainties in quadrature to obtain $\mathrm{[M/H]}=0.27^{+0.14}_{-0.13}$~dex and $^{12}\mathrm{CO}/^{13}\mathrm{CO}=159^{+52}_{-41}$.

Upon comparing the chemical properties from the two fits in Table \ref{tab:comp}, we see that the C/O, metallicity, and carbon isotopologue ratios are all consistent to within $2\sigma$, with the differences being $0.9\sigma$, $1.2\sigma$, and $1.0\sigma$ respectively. In both fits, our retrieved metallicity is consistent with the representative metallicity of objects in AB Doradus, which is very close to solar \citep{Barenfeld_2013}. 

\subsection{Isotopologue Detection and Ratio} \label{subsec:isotope}

To determine whether $^{13}$CO is observed in the brown dwarf, we performed free retrievals including (`full model') and excluding (`reduced model') $^{13}$CO. The two fits yield a $\Delta \ln \mathcal{Z}$ of $4.414$, with the full model being the preferred one, suggesting the presence of $^{13}$CO. 

\begin{figure}[ht!]
    \centering
    \includegraphics[width=\linewidth]{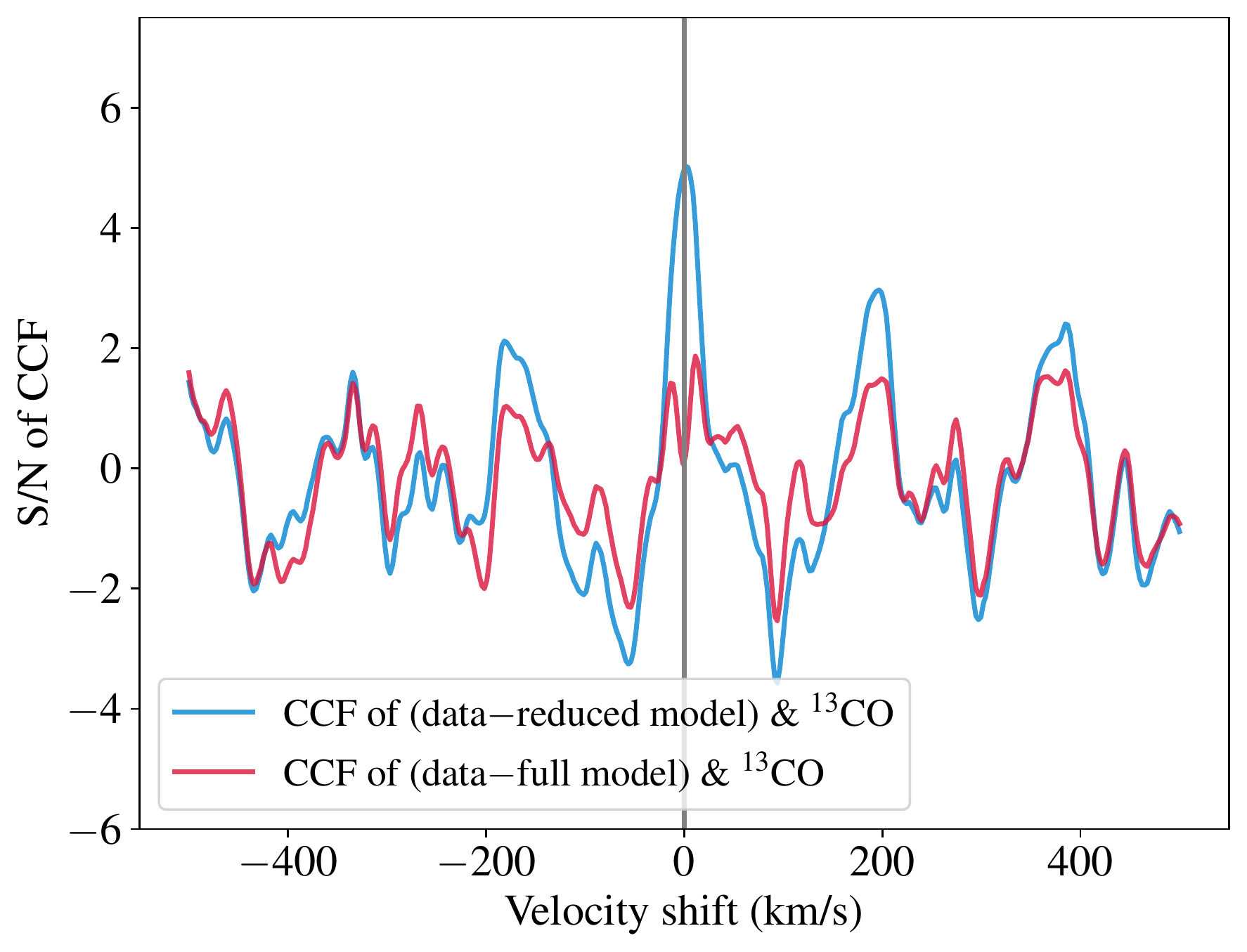}
    \caption{Cross-correlation function for brown dwarf companion with $^{13}$CO template. The `full model' includes $^{13}$CO whereas the `reduced model' does not, and hence residual $^{13}$CO features are only present for the cross-correlation of the data minus the reduced model. The central peak in the blue line indicates a detection of $^{13}$CO. The red line has standard deviation $\sigma = 8.3$~counts, which yields $5.0\sigma$ for the peak in the blue line.}
    \label{fig:ccf13}
\end{figure}

We confirmed this using a cross-correlation function analysis following \citet{xuan2024mdwarf}. First, we subtracted the reduced model from the full model to produce a ``template'' for $^{13}$CO. We then cross-correlated this template with the residuals of each fit of the brown dwarf companion. The two cross-correlation functions from this procedure are shown in Figure \ref{fig:ccf13}, which also indicate a detection. Defining $\sigma$ as the standard deviation of the cross-correlation with the residuals of the full model (which contains no $^{13}$CO), the CCF detection significance is $\approx5.0\sigma$. 

To determine the robustness of the companion's measured isotopologue ratio of $159^{+52}_{-41}$, we constrain the KPIC data's wavelength range to only include the 3-1 transition bandheads for $^{12}$CO and $^{13}$CO, the strongest bandhead available for both isotopologues (see Figure \ref{fig:kpic}). This is motivated by the fact that isotopologue measurements between KPIC and other instruments with different wavelength coverage have sometimes yielded different answers \citep[e.g., GQ Lup B;][]{xuan2024planets, GonzalezPicos2024}. In our case, we find constraining the wavelength range yields values consistent to within $1\sigma$ with our reported ratio. Additionally, if we fix the $^{12}$CO/$^{13}$CO to 70 \citep[comparable to the local interstellar medium;][]{Milam2005}, we retrieve slightly more isothermal but reasonable $P-T$ profiles, in agreement with the inherently large uncertainties on our reported value ($2.2\sigma$ away from 70).

\section{Discussion} \label{sec:discussion}

\subsection{Formation of Brown Dwarf}

Comparing the metallicity of CD-35 2722~B and its host star shows that they are consistent at $1.5\sigma$, and their isotopologue ratios are also consistent at $0.6\sigma$. 
We note there may be biases in our absolute reported isotopologue ratios, but we focus on the consistency between the host and companion. 
This relative agreement suggests that the brown dwarf formed via gravitational instability, through a star-like formation mechanism \citep[e.g.,][]{Balmer_2024, Xuan_2022, xuan2024planets, costes2024, Hsu_2024a}. This is in agreement with our expectations from the brown dwarf's mass and orbital distance. We do not compare the C/O ratio of the two objects because the PHOENIX grid for the star does not include C/O as a free parameter. Furthermore, due to the high temperature of the host star, a significant fraction of the water is expected to be dissociated such that $K$ band spectroscopy alone cannot place meaningful constraints on its C/O ratio \citep{xuan2024mdwarf}.

Recent works have found that lower mass substellar objects tend to have higher metallicities, suggesting a dividing line between populations at $\sim10-15~M_\mathrm{Jup}$ \citep{Zhang_2023, xuan2024planets, Nasedkin_2024, ji_wang2025}. Figure 1 of \citet{ji_wang2025} demonstrates super-stellar metallicities for objects with $\lesssim15~M_\mathrm{Jup}$ and solar metallicites for $\gtrsim20~M_\mathrm{Jup}$ objects, suggesting that the less massive objects accrete more metal-rich solids \citep[e.g., HR8799 c and e;][]{Wang_2023, Molliere_2020}. A caveat is that \citet{ji_wang2025} based their analysis on a limited sample of objects with heterogeneous datasets and model fits, and does not include newer results \citep[e.g.][]{xuan2024planets}. In either case, the nearly solar metallicity we retrieve for an $\approx30~M_\mathrm{Jup}$ brown dwarf agrees with observed trends. 

\subsection{Comparison with AB Dor}

To our knowledge, our $^{12}\mathrm{C}/^{13}\mathrm{C}$ isotopic ratio measurement for CD-35 2722~A is the first for a star in the AB Doradus association. Our measurement for CD-35 2722~B is the fourth for a substellar object in this association, after the L dwarfs 2MASS J03552337+1133437 \citep[$^{12}\mathrm{C}/^{13}\mathrm{C}=97^{+15}_{-11}$;][]{Zhang_2021_2MASS}, 2MASS J03552337+1133437 \citep[$^{12}\mathrm{C}/^{13}\mathrm{C}=96^{+7}_{-6}$;][]{Grasser_2025}, and 2MASS J14252798–3650229 \citep[$^{12}\mathrm{C}/^{13}\mathrm{C}=110^{+11}_{-10}$;][]{Grasser_2025}. 

It will be worthwhile to explore the carbon isotopologue ratio for more objects in this association in the future, which may help place the value we obtain in context. In particular, our measured ratio of $159^{+52}_{-41}$ for CD-35 2722~B is substantially higher than those for the remaining three brown dwarfs, which are consistent with each other to within $1\sigma$. So far, all measurements are higher than the $\sim 68$ for the local interstellar medium \citep{Milam2005}; it may be the case that this is a general trend throughout the association. A larger sample of isotopologue measurements would also help assess the extent of $^{12}$C/$^{13}$C uniformity in AB Dor. 

\subsection{Comparison with Literature Values}

Recently, \citet{Palma-Bifani_2025} studied CD-35 2722~B using medium resolution spectroscopy in the $K$ band, and measured bulk as well as chemical properties. We find general agreement between our radii, C/O ratios, and metallicities. However, \citet{Palma-Bifani_2025} finds a much lower mass of $7.3 \pm 1.1~M_\mathrm{Jup}$, which is in strong tension with both evolutionary predictions of $43 \pm 8~M_\mathrm{Jup}$ as well as our measurement of $30^{+5}_{-4}~M_\mathrm{Jup}$. This suggests the surface gravity they retrieve ($\log g=3.7$ to $4.0$) is too low (evolutionary predictions of $\log g=4.5$ to $5.0$). 

\subsection{Radial Velocities}

The line-of-sight radial velocities of the brown dwarf companion and its host star are $32.99 \pm 0.10$~km/s and $32.28 \pm 0.03$~km/s respectively. Calculating the difference between the two yields a relative radial velocity of $0.71 \pm 0.10$~km/s. We defer updating the brown dwarf's orbit based on this relative RV measurement to future work.

\subsection{Isothermality, Clouds, and Impact on Abundances} \label{subsec:iso}

Even with Gaussian priors on the pressure-temperature structure, we retrieved $P-T$ profiles that were more isothermal than the Diamondback $T_{\mathrm{eff}}=1500$~K model and hotter by about $300$~K in upper parts of the atmosphere above $\sim 10^{-2}$ bar (see Figure \ref{fig:ptp}). We compare to the $T_{\mathrm{eff}}=1500$~K model because it agrees with our retrieved profiles in the deeper pressures from which most of the contribution originates. This phenomenon of being more isothermal than models is a known problem in the literature, especially for cloudy objects, and can result in inaccurate cloud properties and chemical abundances \citep[e.g.][]{Burningham2017, Molliere_2020, Brown-sevilla2023}. 

We do not expect CD-35 2722~B's atmospheric temperature profile to be more isothermal than self-consistent atmospheric models; since it is sufficiently far from its host star, this would require extra heating in the upper atmosphere due to chromospheric activity or accretion \citep{Sorahana_2014}. The typical signature of such accretion is H$\alpha$ emission, which has been found in younger brown dwarfs \citep[e.g., SR12c and Delorme 1 (AB)b;][]{Santamaria_2017, Santamaria_2019, Betti_2022} but is not apparent in our object. Indeed, CD-35 2722~B's $\gtrsim 50$~Myr age is past the typical timescale for accretion of $5-10$~Myr, and in rare cases, up to a few tens of Myr \citep{Zhang_2021, Betti_2022, Luhman2023}. As a result, we also tested an alternative temperature structure treatment where we fixed it to the closest-matching Sonora Diamondback model, finding our retrieved chemical properties were not significantly affected by the increased isothermality compared to models. 

We largely attribute our non-detection of clouds (\S \ref{subsec:noclouds}) to a limitation of $K$ band high-resolution spectroscopy, whose limited wavelength range and normalized continuum are not particularly conducive for detecting clouds. As seen in Figure \ref{fig:ptp}, a large portion of the contribution at our data's wavelengths comes from above the cloud deck pressure, further decreasing our sensitivity to clouds. Our result is similar to the study of $\beta$ Pictoris~b by \citet{Landman2024}, which also did not find clouds in their nominal model with high-resolution data, a result inconsistent with lower-resolution data. We note, however, that clouds have been detected in objects similar to CD-35 2722~B and using similar data as ours, including those from the same instrument \citep[e.g.][]{xuan2024planets}. Observationally, mid-infrared spectra with a wider wavelength coverage would be able to conclusively determine the presence and composition of clouds on CD-35 2722~B \citep[e.g.][]{Luna_2021}. 

We also interpret our results to partially point towards insufficiency in cloud and atmospheric models. Improved modeling of clouds, such as including inhomogeneous, patchy clouds \citep[e.g.][]{Zhang2025} as opposed to the uniform clouds we consider, could provide a more accurate representation of our data. \citet{Luna_2021} also found that cloud models with cloud particles smaller than  those in the EddySed clouds that we used better match \textit{Spitzer} spectra. 

\section{Conclusion} \label{sec:conclusion}

We perform a detailed atmospheric analysis of the L dwarf CD-35 2722~B and its host star, utilizing high-resolution spectroscopy from KPIC and archival photometry. Through equilibrium chemistry retrievals, we constrain the physical and chemical properties of the brown dwarf, shedding light on its formation and atmospheric composition. Adding statistical and systematic uncertainties in quadrature to compare the objects, we measure $\mathrm{[M/H]} = -0.16 \pm 0.25$~dex and $^{12}\mathrm{C}/^{13}\mathrm{C}=132^{+20}_{-14}$ for the host star, and $\mathrm{[M/H]} = 0.27^{+0.14}_{-0.13}$~dex, $^{12}\mathrm{CO}/^{13}\mathrm{CO}=159^{+52}_{-41}$, and $\mathrm{C/O}=0.55\pm0.04$ for the companion. 

For objects such as CD-35 2722~B which are expected to have clouds, there is a known degeneracy between isothermality in the pressure-temperature structure and clouds \citep{Burningham2017, Molliere_2020}. We retrieve slightly isothermal atmospheres compared to self-consistent models, even when placing Gaussian priors on the $P-T$ profile derived from self-consistent models. Fixing the $P-T$ profile to the models removes isothermality and yields chemical properties that are different but consistent when considering systematic uncertainties. We emphasize that our high-resolution $K$ band data are not optimized for cloud detection due to continuum subtraction, a small wavelength range, and much contribution originating from pressures above the cloud deck pressure. This insensitivity to clouds does not significantly impact retrieved chemical properties, but our work demonstrates the need for additional atmospheric modelling as well as data over a wider wavelength range to paint a fuller picture of likely cloudy brown dwarfs. 

\section*{Acknowledgments}

We thank the anonymous reviewer for their feedback, which improved this paper.
We thank Michael Liu and Zahed Wahhaj for helpful discussions on CD-35 2722~B. This work was funded, in part, by a Summer Undergraduate Research Fellowship (SURF) from the California Institute of Technology. J.X. was supported by the NASA Future Investigators in NASA Earth and Space Science and Technology (FINESST) award \#80NSSC23K1434. 

The data presented herein were obtained at Keck Observatory, which is a private 501(c)3 non-profit organization operated as a scientific partnership among the California Institute of Technology, the University of California, and the National Aeronautics and Space Administration. The Observatory was made possible by the generous financial support of the W. M. Keck Foundation. The authors wish to recognize and acknowledge the very significant cultural role and reverence that the summit of Maunakea has always had within the Native Hawaiian community. We are most fortunate to have the opportunity to conduct observations from this mountain.

KPIC is supported by grants from the Heising-Simons Foundation (grants \#2015-129, \#2017-318, \#2019-1312, \#2023-4598), the Simons Foundation, the National Science Foundation (grant AST-1611623), the Jet Propulsion Laboratory, and the California Institute of Technology.

\facilities{Keck:II (KPIC)}

\software{\texttt{petitRADTRANS} \citep{Molliere2019}, \texttt{astropy} \citep{astropy_2013, astropy_2018, astropy_2022}, \texttt{Starfish} \citep{ian_czekala_2018_2221006}, \texttt{dynesty} \citep{Speagle_2020}}

\bibliography{cd35}{}

@ARTICLE{Luhman2023,
       author = {{Luhman}, K.~L. and {Tremblin}, P. and {Birkmann}, S.~M. and {Manjavacas}, E. and {Valenti}, J. and {Alves de Oliveira}, C. and {Beck}, T.~L. and {Giardino}, G. and {L{\"u}tzgendorf}, N. and {Rauscher}, B.~J. and {Sirianni}, M.},
        title = "{JWST/NIRSpec Observations of the Planetary Mass Companion TWA 27B}",
      journal = {\apjl},
         year = 2023,
        month = jun,
       volume = {949},
       number = {2},
          eid = {L36},
        pages = {L36},
          doi = {10.3847/2041-8213/acd635},
archivePrefix = {arXiv},
       eprint = {2305.18603},
 primaryClass = {astro-ph.EP},
       adsurl = {https://ui.adsabs.harvard.edu/abs/2023ApJ...949L..36L}
}

@article{Lee_2024,
doi = {10.3847/1538-4357/ad39e3},
url = {https://dx.doi.org/10.3847/1538-4357/ad39e3},
year = {2024},
month = {jun},
publisher = {The American Astronomical Society},
volume = {969},
number = {1},
pages = {41},
author = {Lee, Seokho and Nomura, Hideko and Furuya, Kenji},
title = {Carbon Isotope Chemistry in Protoplanetary Disks: Effects of C/O Ratios},
journal = {\apj}}

@ARTICLE{Langer_1984,
       author = {{Langer}, W.~D. and {Graedel}, T.~E. and {Frerking}, M.~A. and {Armentrout}, P.~B.},
        title = "{Carbon and oxygen isotope fractionation in dense interstellar clouds.}",
      journal = {\apj},
         year = 1984,
        month = feb,
       volume = {277},
        pages = {581-604},
          doi = {10.1086/161730},
       adsurl = {https://ui.adsabs.harvard.edu/abs/1984ApJ...277..581L}
}

@article{Smith_2015,
doi = {10.1088/0004-637X/813/2/120},
url = {https://dx.doi.org/10.1088/0004-637X/813/2/120},
year = {2015},
month = {nov},
publisher = {The American Astronomical Society},
volume = {813},
number = {2},
pages = {120},
author = {Smith, Rachel L. and Pontoppidan, Klaus M. and Young, Edward D. and Morris, Mark R.},
title = {HETEROGENEITY IN 12CO/13CO ABUNDANCE RATIOS TOWARD SOLAR-TYPE YOUNG STELLAR OBJECTS},
journal = {\apj}}

@ARTICLE{Nasedkin2024,
       author = {{Nasedkin}, Evert and {Molli{\`e}re}, Paul and {Blain}, Doriann},
        title = "{Atmospheric Retrievals with petitRADTRANS}",
      journal = {The Journal of Open Source Software},
         year = 2024,
        month = apr,
       volume = {9},
       number = {96},
          eid = {5875},
        pages = {5875},
          doi = {10.21105/joss.05875},
archivePrefix = {arXiv},
       eprint = {2309.06755},
 primaryClass = {astro-ph.EP},
       adsurl = {https://ui.adsabs.harvard.edu/abs/2024JOSS....9.5875N}
}

@ARTICLE{Jovanovic2025,
       author = {{Jovanovic}, Nemanja and {Echeverri}, Daniel and {Delorme}, Jacques-Robert and {Finnerty}, Luke and {Schofield}, Tobias and {Wang}, Jason J. and {Xin}, Yinzi and {Xuan}, Jerry W. and {Wallace}, J. Kent and {Mawet}, Dimitri and {Sanghi}, Aniket and {Baker}, Ashley and {Bartos}, Randall D. and {Bond}, Charlotte Z. and {Calvin}, Benjamin and {Cetre}, Sylvain and {Doppmann}, Greg and {Fitzgerald}, Michael P. and {Fucik}, Jason R. and {Gao}, Maodong and {Ge}, Jinhao and {Guthery}, Charlotte and {Horstman}, Katelyn and {Hsu}, Chih-Chun and {Liberman}, Joshua and {Leifer}, Stephanie and {Lilley}, Scott and {Lopez}, Ronald A. and {Marin}, Eduardo and {Martin}, Emily C. and {Mennesson}, Bertrand and {Morris}, Evan C. and {Nash}, Reston and {Pezzato}, Jacklyn M. and {Porter}, Michael and {Roberts}, Mitsuko and {Ruane}, Garreth and {Ruffio}, Jean-Baptiste and {Sappey}, Ben and {Serabyn}, Eugene and {Shen}, Boqiang and {Skemer}, Andrew J. and {Wang}, Ji and {Wetherell}, Edward and {Wizinowich}, Peter and {Salama}, Ma{\"\i}ssa and {Chambouleyron}, Vincent and {Jensen-Clem}, Rebecca and {Beichman}, Charles},
        title = "{Technical description and performance of the phase II version of the Keck Planet Imager and Characterizer}",
      journal = {Journal of Astronomical Telescopes, Instruments, and Systems},
         year = 2025,
        month = jan,
       volume = {11},
          eid = {015005},
        pages = {015005},
          doi = {10.1117/1.JATIS.11.1.015005},
archivePrefix = {arXiv},
       eprint = {2502.01863},
 primaryClass = {astro-ph.IM},
       adsurl = {https://ui.adsabs.harvard.edu/abs/2025JATIS..11a5005J}
}

@article{Boss_2023,
doi = {10.3847/1538-4357/acf373},
url = {https://dx.doi.org/10.3847/1538-4357/acf373},
year = {2023},
month = {sep},
publisher = {The American Astronomical Society},
volume = {956},
number = {1},
pages = {4},
author = {Boss, Alan P. and Kanodia, Shubham},
title = {Forming Gas Giants around a Range of Protostellar M-dwarfs by Gas Disk Gravitational Instability},
journal = {\apj}}

@article{Wang_2021, doi = {10.3847/1538-3881/ac1349}, url = {https://dx.doi.org/10.3847/1538-3881/ac1349}, year = {2021}, month = {sep}, publisher = {The American Astronomical Society}, volume = {162}, number = {4}, pages = {148}, author = {Jason J. Wang and Jean-Baptiste Ruffio and Evan Morris and Jacques-Robert Delorme and Nemanja Jovanovic and Jacklyn Pezzato and Daniel Echeverri and Luke Finnerty and Callie Hood and J. J. Zanazzi and Marta L. Bryan and Charlotte Z. Bond and Sylvain Cetre and Emily C. Martin and Dimitri Mawet and Andy Skemer and Ashley Baker and Jerry W. Xuan and J. Kent Wallace and Ji Wang and Randall Bartos and Geoffrey A. Blake and Andy Boden and Cam Buzard and Benjamin Calvin and Mark Chun and Greg Doppmann and Trent J. Dupuy and Gaspard Duchêne and Y. Katherina Feng and Michael P. Fitzgerald and Jonathan Fortney and Richard S. Freedman and Heather Knutson and Quinn Konopacky and Scott Lilley and Michael C. Liu and Ronald Lopez and Roxana Lupu and Mark S. Marley and Tiffany Meshkat and Brittany Miles and Maxwell Millar-Blanchaer and Sam Ragland and Arpita Roy and Garreth Ruane and Ben Sappey and Tobias Schofield and Lauren Weiss and Edward Wetherell and Peter Wizinowich and Marie Ygouf}, title = {Detection and Bulk Properties of the HR 8799 Planets with High-resolution Spectroscopy}, journal = {\aj}}

@article{Zhang_2021, author = {Yapeng Zhang and Ignas A. G. Snellen and Alexander J. Bohn and Paul Mollière and Christian Ginski and H. Jens Hoeijmakers and Matthew A. Kenworthy and Eric E. Mamajek and Tiffany Meshkat and Maddalena Reggiani and Frans Snik}, title = {The $^{13}$CO-rich atmosphere of a young accreting super-Jupiter}, journal = {Nature}, year = {2021}, volume = {595}, number = {7867}, pages = {370--372}, doi = {10.1038/s41586-021-03616-x}, url = {https://doi.org/10.1038/s41586-021-03616-x} }

@article{Xuan_2022, doi = {10.3847/1538-4357/ac8673}, url = {https://dx.doi.org/10.3847/1538-4357/ac8673}, year = {2022}, month = {sep}, publisher = {The American Astronomical Society}, volume = {937}, number = {2}, pages = {54}, author = {Jerry W. Xuan and Jason Wang and Jean-Baptiste Ruffio and Heather Knutson and Dimitri Mawet and Paul Mollière and Jared Kolecki and Arthur Vigan and Sagnick Mukherjee and Nicole Wallack and Ji Wang and Ashley Baker and Randall Bartos and Geoffrey A. Blake and Charlotte Z. Bond and Marta Bryan and Benjamin Calvin and Sylvain Cetre and Mark Chun and Jacques-Robert Delorme and Greg Doppmann and Daniel Echeverri and Luke Finnerty and Michael P. Fitzgerald and Katelyn Horstman and Julie Inglis and Nemanja Jovanovic and Ronald López and Emily C. Martin and Evan Morris and Jacklyn Pezzato and Sam Ragland and Bin Ren and Garreth Ruane and Ben Sappey and Tobias Schofield and Andrew Skemer and Taylor Venenciano and J. Kent Wallace and Peter Wizinowich}, title = {A Clear View of a Cloudy Brown Dwarf Companion from High-resolution Spectroscopy}, journal = {\apj} }

@article{xuan2024planets,
doi = {10.3847/1538-4357/ad4796},
url = {https://dx.doi.org/10.3847/1538-4357/ad4796},
year = {2024},
month = {jul},
publisher = {The American Astronomical Society},
volume = {970},
number = {1},
pages = {71},
author = {Jerry W. Xuan and Chih-Chun Hsu and Luke Finnerty and Jason Wang and Jean-Baptiste Ruffio and Yapeng Zhang and Heather A. Knutson and Dimitri Mawet and Eric E. Mamajek and Julie Inglis and Nicole L. Wallack and Marta L. Bryan and Geoffrey A. Blake and Paul Mollière and Neda Hejazi and Ashley Baker and Randall Bartos and Benjamin Calvin and Sylvain Cetre and Jacques-Robert Delorme and Greg Doppmann and Daniel Echeverri and Michael P. Fitzgerald and Nemanja Jovanovic and Joshua Liberman and Ronald A. López and Evan Morris and Jacklyn Pezzato and Ben Sappey and Tobias Schofield and Andrew Skemer and J. Kent Wallace and Ji Wang and Shubh Agrawal and Katelyn Horstman},
title = {Are These Planets or Brown Dwarfs? Broadly Solar Compositions from High-resolution Atmospheric Retrievals of ∼10–30 MJup Companions},
journal = {\apj}}

@article{xuan2024mdwarf,
    doi = {10.3847/1538-4357/ad1243},
    url = {https://dx.doi.org/10.3847/1538-4357/ad1243},
    year = {2024},
    month = {feb},
    publisher = {The American Astronomical Society},
    volume = {962},
    number = {1},
    pages = {10},
    author = {Jerry W. Xuan and Jason Wang and Luke Finnerty and Katelyn Horstman and Simon Grimm and Anne E. Peck and Eric Nielsen and Heather A. Knutson and Dimitri Mawet and Howard Isaacson and Andrew W. Howard and Michael C. Liu and Sam Walker and Mark W. Phillips and Geoffrey A. Blake and Jean-Baptiste Ruffio and Yapeng Zhang and Julie Inglis and Nicole L. Wallack and Aniket Sanghi and Erica J. Gonzales and Fei Dai and Ashley Baker and Randall Bartos and Charlotte Z. Bond and Marta L. Bryan and Benjamin Calvin and Sylvain Cetre and Jacques-Robert Delorme and Greg Doppmann and Daniel Echeverri and Michael P. Fitzgerald and Nemanja Jovanovic and Joshua Liberman and Ronald A. López and Emily C. Martin and Evan Morris and Jacklyn Pezzato and Garreth Ruane and Ben Sappey and Tobias Schofield and Andrew Skemer and Taylor Venenciano and J. Kent Wallace and Ji Wang and Peter Wizinowich and Yinzi Xin and Shubh Agrawal and Clarissa R. Do Ó and Chih-Chun Hsu and Caprice L. Phillips},
    title = {Validation of Elemental and Isotopic Abundances in Late-M Spectral Types with the Benchmark HIP 55507 AB System},
    journal = {\apj},
}

@article{Wahhaj_2011, doi = {10.1088/0004-637X/729/2/139}, url = {https://dx.doi.org/10.1088/0004-637X/729/2/139}, year = {2011}, month = {feb}, publisher = {The American Astronomical Society}, volume = {729}, number = {2}, pages = {139}, author = {Zahed Wahhaj and Michael C. Liu and Beth A. Biller and Fraser Clarke and Eric L. Nielsen and Laird M. Close and Thomas L. Hayward and Eric E. Mamajek and Michael Cushing and Trent Dupuy and Matthias Tecza and Niranjan Thatte and Mark Chun and Christ Ftaclas and Markus Hartung and I. Neill Reid and Evgenya L. Shkolnik and Silvia H. P. Alencar and Pawel Artymowicz and Alan Boss and Elisabethe de Gouveia Dal Pino and Jane Gregorio-Hetem and Shigeru Ida and Marc Kuchner and Douglas N. C. Lin and Douglas W. Toomey}, title = {THE GEMINI NICI PLANET-FINDING CAMPAIGN: DISCOVERY OF A SUBSTELLAR L DWARF COMPANION TO THE NEARBY YOUNG M DWARF CD−35 2722*}, journal = {\apj} }

@article{Delorme2021, author = {Jacques-Robert Delorme and Nemanja Jovanovic and Daniel Echeverri and Dimitri P. Mawet and James Kent Wallace and Randall D. Bartos and Sylvain Cetre and Peter L. Wizinowich and Sam Ragland and Scott J. Lilley and Edward Wetherell and Greg Doppmann and Jason J. Wang and Evan C. Morris and Jean-Baptiste Ruffio and Emily C. Martin and Michael P. Fitzgerald and Garreth J. Ruane and Tobias Schofield and Nick Suominen and Benjamin Calvin and Eric Wang and Kenneth G. Magnone and Christopher A. Johnson and Ji Man Sohn and Ronald A. Lopez and Charlotte Z. Bond and Jacklyn Pezzato and Jorge Llop-Sayson and Mark R. Chun and Andrew J. Skemer}, title = {{Keck Planet Imager and Characterizer: a dedicated single-mode fiber injection unit for high-resolution exoplanet spectroscopy}}, volume = {7}, journal = {Journal of Astronomical Telescopes, Instruments, and Systems}, number = {3}, publisher = {SPIE}, pages = {035006}, year = {2021}, doi = {10.1117/1.JATIS.7.3.035006}, URL = {https://doi.org/10.1117/1.JATIS.7.3.035006} }

@article{Mawet_2016, author = {D. Mawet and P. Wizinowich and R. Dekany and M. Chun and D. Hall and S. Cetre and O. Guyon and J. K. Wallace and B. Bowler and M. Liu and G. Ruane and E. Serabyn and R. Bartos and J. Wang and G. Vasisht and M. Fitzgerald and A. Skemer and M. Ireland and J. Fucik and J. Fortney and I. Crossfield and R. Hu and B. Benneke}, title = {{Keck Planet Imager and Characterizer: concept and phased implementation}}, volume = {9909}, journal = {Adaptive Optics Systems V}, editor = {Enrico Marchetti and Laird M. Close and Jean-Pierre Véran}, organization = {International Society for Optics and Photonics}, publisher = {SPIE}, pages = {99090D}, year = {2016}, doi = {10.1117/12.2233658}, URL = {https://doi.org/10.1117/12.2233658} }

@article{Molliere2019, author = {Mollière, P. and Wardenier, J. P. and van Boekel, R. and Henning, Th. and Molaverdikhani, K. and Snellen, I. A. G.}, title = {petitRADTRANS - A Python radiative transfer package for exoplanet characterization and retrieval}, DOI = "10.1051/0004-6361/201935470", url = "https://doi.org/10.1051/0004-6361/201935470", journal = {A\&A}, year = 2019, volume = 627, pages = "A67" }

@article{morley2024sonora,
doi = {10.3847/1538-4357/ad71d5},
url = {https://dx.doi.org/10.3847/1538-4357/ad71d5},
year = {2024},
month = {oct},
publisher = {The American Astronomical Society},
volume = {975},
number = {1},
pages = {59},
author = {Caroline V. Morley and Sagnick Mukherjee and Mark S. Marley and Jonathan J. Fortney and Channon Visscher and Roxana Lupu and Ehsan Gharib-Nezhad and Daniel Thorngren and Richard Freedman and Natasha Batalha},
title = {The Sonora Substellar Atmosphere Models. III. Diamondback: Atmospheric Properties, Spectra, and Evolution for Warm Cloudy Substellar Objects},
journal = {\apj}}

@article{Nasedkin_2024,
	author = {{Nasedkin, E.} and {Mollière, P.} and {Lacour, S.} and {Nowak, M.} and {Kreidberg, L.} and {Stolker, T.} and {Wang, J. J.} and {Balmer, W. O.} and {Kammerer, J.} and {Shangguan, J.} and {Abuter, R.} and {Amorim, A.} and {Asensio-Torres, R.} and {Benisty, M.} and {Berger, J.-P.} and {Beust, H.} and {Blunt, S.} and {Boccaletti, A.} and {Bonnefoy, M.} and {Bonnet, H.} and {Bordoni, M. S.} and {Bourdarot, G.} and {Brandner, W.} and {Cantalloube, F.} and {Caselli, P.} and {Charnay, B.} and {Chauvin, G.} and {Chavez, A.} and {Choquet, E.} and {Christiaens, V.} and {Clénet, Y.} and {Coudé du Foresto, V.} and {Cridland, A.} and {Davies, R.} and {Dembet, R.} and {Dexter, J.} and {Drescher, A.} and {Duvert, G.} and {Eckart, A.} and {Eisenhauer, F.} and {Förster Schreiber, N. M.} and {Garcia, P.} and {Garcia Lopez, R.} and {Gendron, E.} and {Genzel, R.} and {Gillessen, S.} and {Girard, J. H.} and {Grant, S.} and {Haubois, X.} and {Heißel, G.} and {Henning, Th.} and {Hinkley, S.} and {Hippler, S.} and {Houllé, M.} and {Hubert, Z.} and {Jocou, L.} and {Keppler, M.} and {Kervella, P.} and {Kurtovic, N. T.} and {Lagrange, A.-M.} and {Lapeyrère, V.} and {Le Bouquin, J.-B.} and {Lutz, D.} and {Maire, A.-L.} and {Mang, F.} and {Marleau, G.-D.} and {Mérand, A.} and {Monnier, J. D.} and {Mordasini, C.} and {Ott, T.} and {Otten, G. P. P. L.} and {Paladini, C.} and {Paumard, T.} and {Perraut, K.} and {Perrin, G.} and {Pfuhl, O.} and {Pourré, N.} and {Pueyo, L.} and {Ribeiro, D. C.} and {Rickman, E.} and {Ruffio, J. B.} and {Rustamkulov, Z.} and {Shimizu, T.} and {Sing, D.} and {Stadler, J.} and {Straub, O.} and {Straubmeier, C.} and {Sturm, E.} and {Tacconi, L. J.} and {van Dishoeck, E. F.} and {Vigan, A.} and {Vincent, F.} and {von Fellenberg, S. D.} and {Widmann, F.} and {Winterhalder, T. O.} and {Woillez, J.} and {Yazici, Ş.} and {the GRAVITY Collaboration}},
	title = {Four-of-a-kind? Comprehensive atmospheric characterisation of the HR 8799 planets with VLTI/GRAVITY},
	DOI= "10.1051/0004-6361/202449328",
	url= "https://doi.org/10.1051/0004-6361/202449328",
	journal = {A\&A},
	year = 2024,
	volume = 687,
	pages = "A298",
}

@article{Zhang_2024,
doi = {10.3847/1538-3881/ad7ea9},
url = {https://dx.doi.org/10.3847/1538-3881/ad7ea9},
year = {2024},
month = {nov},
publisher = {The American Astronomical Society},
volume = {168},
number = {6},
pages = {246},
author = {Zhang, Yapeng and González Picos, Darío and de Regt, Sam and Snellen, Ignas A. G. and Gandhi, Siddharth and Ginski, Christian and Kesseli, Aurora Y. and Landman, Rico and Mollière, Paul and Nasedkin, Evert and Sánchez-López, Alejandro and Stolker, Tomas and Inglis, Julie and Knutson, Heather A. and Mawet, Dimitri and Wallack, Nicole and Xuan, Jerry W.},
title = {The ESO SupJup Survey. III. Confirmation of 13CO in YSES 1 b and Atmospheric Detection of YSES 1 c with CRIRES+},
journal = {\aj}}

@article{Marley_2021, doi = {10.3847/1538-4357/ac141d}, url = {https://dx.doi.org/10.3847/1538-4357/ac141d}, year = {2021}, month = {oct}, publisher = {The American Astronomical Society}, volume = {920}, number = {2}, pages = {85}, author = {Mark S. Marley and Didier Saumon and Channon Visscher and Roxana Lupu and Richard Freedman and Caroline Morley and Jonathan J. Fortney and Christopher Seay and Adam J. R. W. Smith and D. J. Teal and Ruoyan Wang}, title = {The Sonora Brown Dwarf Atmosphere and Evolution Models. I. Model Description and Application to Cloudless Atmospheres in Rainout Chemical Equilibrium}, journal = {\apj} }

@article{Schlaufman_2018,
doi = {10.3847/1538-4357/aa961c},
url = {https://dx.doi.org/10.3847/1538-4357/aa961c},
year = {2018},
month = {jan},
publisher = {The American Astronomical Society},
volume = {853},
number = {1},
pages = {37},
author = {Kevin C. Schlaufman},
title = {Evidence of an Upper Bound on the Masses of Planets and Its Implications for Giant Planet Formation},
journal = {\apj}}

@article{Spiegel_2011,
doi = {10.1088/0004-637X/727/1/57},
url = {https://dx.doi.org/10.1088/0004-637X/727/1/57},
year = {2011},
month = {jan},
publisher = {The American Astronomical Society},
volume = {727},
number = {1},
pages = {57},
author = {David S. Spiegel and Adam Burrows and John A. Milsom},
title = {THE DEUTERIUM-BURNING MASS LIMIT FOR BROWN DWARFS AND GIANT PLANETS},
journal = {\apj}}

@article{Dieterich_2014,
doi = {10.1088/0004-6256/147/5/94},
url = {https://dx.doi.org/10.1088/0004-6256/147/5/94},
year = {2014},
month = {mar},
publisher = {The American Astronomical Society},
volume = {147},
number = {5},
pages = {94},
author = {Sergio B. Dieterich and Todd J. Henry and Wei-Chun Jao and Jennifer G. Winters and Altonio D. Hosey and Adric R. Riedel and John P. Subasavage},
title = {THE SOLAR NEIGHBORHOOD. XXXII. THE HYDROGEN BURNING LIMIT*,†},
journal = {\aj}}

@article{Burrows_2001,
  title = {The theory of brown dwarfs and extrasolar giant planets},
  author = {Burrows, Adam and Hubbard, W. B. and Lunine, J. I. and Liebert, James},
  journal = {Rev. Mod. Phys.},
  volume = {73},
  issue = {3},
  pages = {719--765},
  numpages = {0},
  year = {2001},
  month = {Sep},
  publisher = {American Physical Society},
  doi = {10.1103/RevModPhys.73.719},
  url = {https://link.aps.org/doi/10.1103/RevModPhys.73.719}
}

@article{Calamari_2024,
doi = {10.3847/1538-4357/ad1f6d},
url = {https://dx.doi.org/10.3847/1538-4357/ad1f6d},
year = {2024},
month = {feb},
publisher = {The American Astronomical Society},
volume = {963},
number = {1},
pages = {67},
author = {Emily Calamari and Jacqueline K. Faherty and Channon Visscher and Marina E. Gemma and Ben Burningham and Austin Rothermich},
title = {Predicting Cloud Conditions in Substellar Mass Objects Using Ultracool Dwarf Companions},
journal = {\apj}}

@article{Palma-Bifani_2025,
	author = {{Palma-Bifani, P.} and {Bonnefoy, M.} and {Chauvin, G.} and {Rojo, P.} and {Baudoz, P.} and {Charnay, B.} and {Denis, A.} and {Hoch, K.} and {Petrus, S.} and {Ravet, M.} and {Simonnin, A.} and {Vigan, A.}},
	title = {The planetary-mass-limit VLT/SINFONI library - Spectral extraction and atmospheric characterization via forward modeling★},
	DOI= "10.1051/0004-6361/202554894",
	url= "https://doi.org/10.1051/0004-6361/202554894",
	journal = {A\&A},
	year = 2025,
	volume = 701,
	pages = "A51",
}

@article{piette2020, author = {Piette, Anjali A A and Madhusudhan, Nikku}, title = "{Considerations for atmospheric retrieval of high-precision brown dwarf spectra}", journal = {\mnras}, volume = {497}, number = {4}, pages = {5136-5154}, year = {2020}, month = {08}, issn = {0035-8711}, doi = {10.1093/mnras/staa2289}, url = {https://doi.org/10.1093/mnras/staa2289}, eprint = {https://academic.oup.com/mnras/article-pdf/497/4/5136/33687754/staa2289.pdf} }

@article{Ackerman_2001, doi = {10.1086/321540}, url = {https://dx.doi.org/10.1086/321540}, year = {2001}, month = {aug}, publisher = {}, volume = {556}, number = {2}, pages = {872}, author = {Andrew S. Ackerman and Mark S. Marley}, title = {Precipitating Condensation Clouds in Substellar Atmospheres}, journal = {\apj} }

@article{Min_2005, author = {Min, M. and {Hovenier, J. W.} and {de Koter, A.}}, title = {Modeling optical properties of cosmic dust grains using a distribution of hollow spheres}, DOI = "10.1051/0004-6361:20041920", url = "https://doi.org/10.1051/0004-6361:20041920", journal = {A\&A}, year = 2005, volume = 432, number = 3, pages = "909-920" }

@article{Burningham2017, author = {Burningham, Ben and Marley, M. S. and Line, M. R. and Lupu, R. and Visscher, C. and Morley, C. V. and Saumon, D. and Freedman, R.}, title = "{Retrieval of atmospheric properties of cloudy L dwarfs}", journal = {\mnras}, volume = {470}, number = {1}, pages = {1177-1197}, year = {2017}, month = {05}, issn = {0035-8711}, doi = {10.1093/mnras/stx1246}, url = {https://doi.org/10.1093/mnras/stx1246}, eprint = {https://academic.oup.com/mnras/article-pdf/470/1/1177/17933870/stx1246.pdf} }

@article{Brown-sevilla2023, author = {{Brown-Sevilla, S. B.} and {Maire, A.-L.} and {Mollière, P.} and {Samland, M.} and {Feldt, M.} and {Brandner, W.} and {Henning, Th.} and {Gratton, R.} and {Janson, M.} and {Stolker, T.} and {Hagelberg, J.} and {Zurlo, A.} and {Cantalloube, F.} and {Boccaletti, A.} and {Bonnefoy, M.} and {Chauvin, G.} and {Desidera, S.} and {D'Orazi, V.} and {Lagrange, A.-M.} and {Langlois, M.} and {Menard, F.} and {Mesa, D.} and {Meyer, M.} and {Pavlov, A.} and {Petit, C.} and {Rochat, S.} and {Rouan, D.} and {Schmidt, T.} and {Vigan, A.} and {Weber, L.}}, title = {Revisiting the atmosphere of the exoplanet 51 Eridani b with VLT/SPHERE★}, DOI = "10.1051/0004-6361/202244826", url = "https://doi.org/10.1051/0004-6361/202244826", journal = {A\&A}, year = 2023, volume = 673, pages = "A98" }

@INCOLLECTION{Marley_2013, author = {{Marley}, M.~S. and {Ackerman}, A.~S. and {Cuzzi}, J.~N. and {Kitzmann}, D.}, title = "{Clouds and Hazes in Exoplanet Atmospheres}", booktitle = {Comparative Climatology of Terrestrial Planets}, year = 2013, editor = {{Mackwell}, Stephen J. and {Simon-Miller}, Amy A. and {Harder}, Jerald W. and {Bullock}, Mark A.}, pages = {367-392}, doi = {10.2458/azu_uapress_9780816530595-ch015}, publisher = {University of Arizona Press}, adsurl = {https://ui.adsabs.harvard.edu/abs/2013cctp.book..367M} }

@article{Speagle_2020, author = {Speagle, Joshua S}, title = "{dynesty: a dynamic nested sampling package for estimating Bayesian posteriors and evidences}", journal = {\mnras}, volume = {493}, number = {3}, pages = {3132-3158}, year = {2020}, month = {02}, issn = {0035-8711}, doi = {10.1093/mnras/staa278}, url = {https://doi.org/10.1093/mnras/staa278}, eprint = {https://academic.oup.com/mnras/article-pdf/493/3/3132/32890730/staa278.pdf} }

@article{Bell_2015,
    author = {Bell, Cameron P. M. and Mamajek, Eric E. and Naylor, Tim},
    title = "{A self-consistent, absolute isochronal age scale for young moving groups in the solar neighbourhood}",
    journal = {\mnras},
    volume = {454},
    number = {1},
    pages = {593-614},
    year = {2015},
    month = {09},
    issn = {0035-8711},
    doi = {10.1093/mnras/stv1981},
    url = {https://doi.org/10.1093/mnras/stv1981},
    eprint = {https://academic.oup.com/mnras/article-pdf/454/1/593/13769516/stv1981.pdf},
}

@article{Elliott_2016,
	author = {Elliott, P. and {Bayo, A.} and {Melo, C. H. F.} and {Torres, C. A. O.} and {Sterzik, M. F.} and {Quast, G. R.} and {Montes, D.} and {Brahm, R.}},
	title = {Search for associations containing young stars (SACY) - VII. New stellar and substellar candidate members in the young associations},
	DOI= "10.1051/0004-6361/201628253",
	url= "https://doi.org/10.1051/0004-6361/201628253",
	journal = {A\&A},
	year = 2016,
	volume = 590,
	pages = "A13",
}

@article{Gagne_2014,
doi = {10.1088/0004-637X/783/2/121},
url = {https://dx.doi.org/10.1088/0004-637X/783/2/121},
year = {2014},
month = {feb},
publisher = {The American Astronomical Society},
volume = {783},
number = {2},
pages = {121},
author = {Jonathan Gagné and David Lafrenière and René Doyon and Lison Malo and Étienne Artigau},
title = {BANYAN. II. VERY LOW MASS AND SUBSTELLAR CANDIDATE MEMBERS TO NEARBY, YOUNG KINEMATIC GROUPS WITH PREVIOUSLY KNOWN SIGNS OF YOUTH},
journal = {\apj}}

@article{Chabrier_2023,
	author = {Chabrier, Gilles and Baraffe, Isabelle and Phillips, Mark and Debras, Florian},
	title = {Impact of a new H/He equation of state on the evolution of massive brown dwarfs - New determination of the hydrogen burning limit},
	DOI= "10.1051/0004-6361/202243832",
	url= "https://doi.org/10.1051/0004-6361/202243832",
	journal = {A\&A},
	year = 2023,
	volume = 671,
	pages = "A119",
}

@article{Molliere_2020,
	author = {{Mollière, P.} and {Stolker, T.} and {Lacour, S.} and {Otten, G. P. P. L.} and {Shangguan, J.} and {Charnay, B.} and {Molyarova, T.} and {Nowak, M.} and {Henning, Th.} and {Marleau, G.-D.} and {Semenov, D. A.} and {van Dishoeck, E.} and {Eisenhauer, F.} and {Garcia, P.} and {Garcia Lopez, R.} and {Girard, J. H.} and {Greenbaum, A. Z.} and {Hinkley, S.} and {Kervella, P.} and {Kreidberg, L.} and {Maire, A.-L.} and {Nasedkin, E.} and {Pueyo, L.} and {Snellen, I. A. G.} and {Vigan, A.} and {Wang, J.} and {de Zeeuw, P. T.} and {Zurlo, A.}},
	title = {Retrieving scattering clouds and disequilibrium chemistry in the atmosphere of HR 8799e},
	DOI= "10.1051/0004-6361/202038325",
	url= "https://doi.org/10.1051/0004-6361/202038325",
	journal = {A\&A},
	year = 2020,
	volume = 640,
	pages = "A131",
}

@article{Luna_2021,
doi = {10.3847/1538-4357/ac1865},
url = {https://dx.doi.org/10.3847/1538-4357/ac1865},
year = {2021},
month = {oct},
publisher = {The American Astronomical Society},
volume = {920},
number = {2},
pages = {146},
author = {Jessica L. Luna and Caroline V. Morley},
title = {Empirically Determining Substellar Cloud Compositions in the Era of the James Webb Space Telescope},
journal = {\apj}}

@article{Martin_2018,
author = {Emily C. Martin and Michael P. Fitzgerald and Ian S. McLean and Gregory Doppmann and Marc Kassis and Ted Aliado and John Canfield and Chris Johnson and Evan Kress and Kyle Lanclos and Kenneth Magnone and Ji Man Sohn and Eric Wang and Jason Weiss},
title = {{An overview of the NIRSPEC upgrade for the Keck II telescope}},
volume = {10702},
booktitle = {Ground-based and Airborne Instrumentation for Astronomy VII},
editor = {Christopher J. Evans and Luc Simard and Hideki Takami},
organization = {International Society for Optics and Photonics},
publisher = {SPIE},
pages = {107020A},
year = {2018},
doi = {10.1117/12.2312266},
URL = {https://doi.org/10.1117/12.2312266}
}

@article{Lopez_2020,
author = {Ronald A. L{\'o}pez and Erika B. Hoffman and Greg Doppmann and Michael P. Fitzgerald and Chris Johnson and Marc Kassis and Kyle Lanclos and Jim Lyke and Emily C. Martin and Ian McLean and Ji Man Sohn and Jason Weiss},
title = {{Characterization and performance of the upgraded NIRSPEC on the W. M. Keck Telescope}},
volume = {11447},
booktitle = {Ground-based and Airborne Instrumentation for Astronomy VIII},
editor = {Christopher J. Evans and Julia J. Bryant and Kentaro Motohara},
organization = {International Society for Optics and Photonics},
publisher = {SPIE},
pages = {114476B},
year = {2020},
doi = {10.1117/12.2563075},
URL = {https://doi.org/10.1117/12.2563075}
}

@article{Saumon_2008,
doi = {10.1086/592734},
url = {https://dx.doi.org/10.1086/592734},
year = {2008},
month = {dec},
publisher = {},
volume = {689},
number = {2},
pages = {1327},
author = {D. Saumon and Mark S. Marley},
title = {The Evolution of L and T Dwarfs in Color-Magnitude Diagrams},
journal = {\apj}}

@article{Phillips_2020,
	author = {Phillips, M. W. and Tremblin, P. and Baraffe, I. and Chabrier, G. and Allard, N. F. and Spiegelman, F. and Goyal, J. M. and Drummond, B. and Hébrard, E.},
	title = {A new set of atmosphere and evolution models for cool T–Y brown dwarfs and giant exoplanets},
	DOI= "10.1051/0004-6361/201937381",
	url= "https://doi.org/10.1051/0004-6361/201937381",
	journal = {A\&A},
	year = 2020,
	volume = 637,
	pages = "A38",
}

@article{Allard_2001,
doi = {10.1086/321547},
url = {https://dx.doi.org/10.1086/321547},
year = {2001},
month = {jul},
publisher = {},
volume = {556},
number = {1},
pages = {357},
author = {France Allard and Peter H. Hauschildt and David R. Alexander and Akemi Tamanai and Andreas Schweitzer},
title = {The Limiting Effects of Dust in Brown Dwarf Model Atmospheres},
journal = {\apj}}

@article{Oberg_2011,
doi = {10.1088/2041-8205/743/1/L16},
url = {https://dx.doi.org/10.1088/2041-8205/743/1/L16},
year = {2011},
month = {nov},
publisher = {The American Astronomical Society},
volume = {743},
number = {1},
pages = {L16},
author = {Karin I. Öberg and Ruth Murray-Clay and Edwin A. Bergin},
title = {THE EFFECTS OF SNOWLINES ON C/O IN PLANETARY ATMOSPHERES},
journal = {\apjl}}

@article{Zhang_2023,
doi = {10.3847/1538-3881/acf768},
url = {https://dx.doi.org/10.3847/1538-3881/acf768},
year = {2023},
month = {oct},
publisher = {The American Astronomical Society},
volume = {166},
number = {5},
pages = {198},
author = {Zhoujian Zhang and Paul Mollière and Keith Hawkins and Catherine Manea and Jonathan J. Fortney and Caroline V. Morley and Andrew Skemer and Mark S. Marley and Brendan P. Bowler and Aarynn L. Carter and Kyle Franson and Zachary G. Maas and Christopher Sneden},journal = {\aj}, title = {ELemental abundances of Planets and brown dwarfs Imaged around Stars (ELPIS). I. Potential Metal Enrichment of the Exoplanet AF Lep b and a Novel Retrieval Approach for Cloudy Self-luminous Atmospheres}}

@article{Gaia_2021,
	author = {{Gaia Collaboration} and Brown, A. G. A. and {Vallenari, A.} and {Prusti, T.} and {de Bruijne, J. H. J.} and {Babusiaux, C.} and {Biermann, M.} and {Creevey, O. L.} and {Evans, D. W.} and {Eyer, L.} and {Hutton, A.} and {Jansen, F.} and {Jordi, C.} and {Klioner, S. A.} and {Lammers, U.} and {Lindegren, L.} and {Luri, X.} and {Mignard, F.} and {Panem, C.} and {Pourbaix, D.} and {Randich, S.} and {Sartoretti, P.} and {Soubiran, C.} and {Walton, N. A.} and {Arenou, F.} and {Bailer-Jones, C. A. L.} and {Bastian, U.} and {Cropper, M.} and {Drimmel, R.} and {Katz, D.} and {Lattanzi, M. G.} and {van Leeuwen, F.} and {Bakker, J.} and {Cacciari, C.} and {Castañeda, J.} and {De Angeli, F.} and {Ducourant, C.} and {Fabricius, C.} and {Fouesneau, M.} and {Frémat, Y.} and {Guerra, R.} and {Guerrier, A.} and {Guiraud, J.} and {Jean-Antoine Piccolo, A.} and {Masana, E.} and {Messineo, R.} and {Mowlavi, N.} and {Nicolas, C.} and {Nienartowicz, K.} and {Pailler, F.} and {Panuzzo, P.} and {Riclet, F.} and {Roux, W.} and {Seabroke, G. M.} and {Sordo, R.} and {Tanga, P.} and {Thévenin, F.} and {Gracia-Abril, G.} and {Portell, J.} and {Teyssier, D.} and {Altmann, M.} and {Andrae, R.} and {Bellas-Velidis, I.} and {Benson, K.} and {Berthier, J.} and {Blomme, R.} and {Brugaletta, E.} and {Burgess, P. W.} and {Busso, G.} and {Carry, B.} and {Cellino, A.} and {Cheek, N.} and {Clementini, G.} and {Damerdji, Y.} and {Davidson, M.} and {Delchambre, L.} and {Dell’Oro, A.} and {Fernández-Hernández, J.} and {Galluccio, L.} and {García-Lario, P.} and {Garcia-Reinaldos, M.} and {González-Núñez, J.} and {Gosset, E.} and {Haigron, R.} and {Halbwachs, J.-L.} and {Hambly, N. C.} and {Harrison, D. L.} and {Hatzidimitriou, D.} and {Heiter, U.} and {Hernández, J.} and {Hestroffer, D.} and {Hodgkin, S. T.} and {Holl, B.} and {Janßen, K.} and {Jevardat de Fombelle, G.} and {Jordan, S.} and {Krone-Martins, A.} and {Lanzafame, A. C.} and {Löffler, W.} and {Lorca, A.} and {Manteiga, M.} and {Marchal, O.} and {Marrese, P. M.} and {Moitinho, A.} and {Mora, A.} and {Muinonen, K.} and {Osborne, P.} and {Pancino, E.} and {Pauwels, T.} and {Petit, J.-M.} and {Recio-Blanco, A.} and {Richards, P. J.} and {Riello, M.} and {Rimoldini, L.} and {Robin, A. C.} and {Roegiers, T.} and {Rybizki, J.} and {Sarro, L. M.} and {Siopis, C.} and {Smith, M.} and {Sozzetti, A.} and {Ulla, A.} and {Utrilla, E.} and {van Leeuwen, M.} and {van Reeven, W.} and {Abbas, U.} and {Abreu Aramburu, A.} and {Accart, S.} and {Aerts, C.} and {Aguado, J. J.} and {Ajaj, M.} and {Altavilla, G.} and {Álvarez, M. A.} and {Álvarez Cid-Fuentes, J.} and {Alves, J.} and {Anderson, R. I.} and {Anglada Varela, E.} and {Antoja, T.} and {Audard, M.} and {Baines, D.} and {Baker, S. G.} and {Balaguer-Núñez, L.} and {Balbinot, E.} and {Balog, Z.} and {Barache, C.} and {Barbato, D.} and {Barros, M.} and {Barstow, M. A.} and {Bartolomé, S.} and {Bassilana, J.-L.} and {Bauchet, N.} and {Baudesson-Stella, A.} and {Becciani, U.} and {Bellazzini, M.} and {Bernet, M.} and {Bertone, S.} and {Bianchi, L.} and {Blanco-Cuaresma, S.} and {Boch, T.} and {Bombrun, A.} and {Bossini, D.} and {Bouquillon, S.} and {Bragaglia, A.} and {Bramante, L.} and {Breedt, E.} and {Bressan, A.} and {Brouillet, N.} and {Bucciarelli, B.} and {Burlacu, A.} and {Busonero, D.} and {Butkevich, A. G.} and {Buzzi, R.} and {Caffau, E.} and {Cancelliere, R.} and {Cánovas, H.} and {Cantat-Gaudin, T.} and {Carballo, R.} and {Carlucci, T.} and {Carnerero, M. I} and {Carrasco, J. M.} and {Casamiquela, L.} and {Castellani, M.} and {Castro-Ginard, A.} and {Castro Sampol, P.} and {Chaoul, L.} and {Charlot, P.} and {Chemin, L.} and {Chiavassa, A.} and {Cioni, M.-R. L.} and {Comoretto, G.} and {Cooper, W. J.} and {Cornez, T.} and {Cowell, S.} and {Crifo, F.} and {Crosta, M.} and {Crowley, C.} and {Dafonte, C.} and {Dapergolas, A.} and {David, M.} and {David, P.} and {de Laverny, P.} and {De Luise, F.} and {De March, R.} and {De Ridder, J.} and {de Souza, R.} and {de Teodoro, P.} and {de Torres, A.} and {del Peloso, E. F.} and {del Pozo, E.} and {Delbo, M.} and {Delgado, A.} and {Delgado, H. E.} and {Delisle, J.-B.} and {Di Matteo, P.} and {Diakite, S.} and {Diener, C.} and {Distefano, E.} and {Dolding, C.} and {Eappachen, D.} and {Edvardsson, B.} and {Enke, H.} and {Esquej, P.} and {Fabre, C.} and {Fabrizio, M.} and {Faigler, S.} and {Fedorets, G.} and {Fernique, P.} and {Fienga, A.} and {Figueras, F.} and {Fouron, C.} and {Fragkoudi, F.} and {Fraile, E.} and {Franke, F.} and {Gai, M.} and {Garabato, D.} and {Garcia-Gutierrez, A.} and {García-Torres, M.} and {Garofalo, A.} and {Gavras, P.} and {Gerlach, E.} and {Geyer, R.} and {Giacobbe, P.} and {Gilmore, G.} and {Girona, S.} and {Giuffrida, G.} and {Gomel, R.} and {Gomez, A.} and {Gonzalez-Santamaria, I.} and {González-Vidal, J. J.} and {Granvik, M.} and {Gutiérrez-Sánchez, R.} and {Guy, L. P.} and {Hauser, M.} and {Haywood, M.} and {Helmi, A.} and {Hidalgo, S. L.} and {Hilger, T.} and {Hładczuk, N.} and {Hobbs, D.} and {Holland, G.} and {Huckle, H. E.} and {Jasniewicz, G.} and {Jonker, P. G.} and {Juaristi Campillo, J.} and {Julbe, F.} and {Karbevska, L.} and {Kervella, P.} and {Khanna, S.} and {Kochoska, A.} and {Kontizas, M.} and {Kordopatis, G.} and {Korn, A. J.} and {Kostrzewa-Rutkowska, Z.} and {Kruszyńska, K.} and {Lambert, S.} and {Lanza, A. F.} and {Lasne, Y.} and {Le Campion, J.-F.} and {Le Fustec, Y.} and {Lebreton, Y.} and {Lebzelter, T.} and {Leccia, S.} and {Leclerc, N.} and {Lecoeur-Taibi, I.} and {Liao, S.} and {Licata, E.} and {Lindstrøm, E. P.} and {Lister, T. A.} and {Livanou, E.} and {Lobel, A.} and {Madrero Pardo, P.} and {Managau, S.} and {Mann, R. G.} and {Marchant, J. M.} and {Marconi, M.} and {Marcos Santos, M. M. S.} and {Marinoni, S.} and {Marocco, F.} and {Marshall, D. J.} and {Martin Polo, L.} and {Martín-Fleitas, J. M.} and {Masip, A.} and {Massari, D.} and {Mastrobuono-Battisti, A.} and {Mazeh, T.} and {McMillan, P. J.} and {Messina, S.} and {Michalik, D.} and {Millar, N. R.} and {Mints, A.} and {Molina, D.} and {Molinaro, R.} and {Molnár, L.} and {Montegriffo, P.} and {Mor, R.} and {Morbidelli, R.} and {Morel, T.} and {Morris, D.} and {Mulone, A. F.} and {Munoz, D.} and {Muraveva, T.} and {Murphy, C. P.} and {Musella, I.} and {Noval, L.} and {Ordénovic, C.} and {Orrù, G.} and {Osinde, J.} and {Pagani, C.} and {Pagano, I.} and {Palaversa, L.} and {Palicio, P. A.} and {Panahi, A.} and {Pawlak, M.} and {Peñalosa Esteller, X.} and {Penttilä, A.} and {Piersimoni, A. M.} and {Pineau, F.-X.} and {Plachy, E.} and {Plum, G.} and {Poggio, E.} and {Poretti, E.} and {Poujoulet, E.} and {Prša, A.} and {Pulone, L.} and {Racero, E.} and {Ragaini, S.} and {Rainer, M.} and {Raiteri, C. M.} and {Rambaux, N.} and {Ramos, P.} and {Ramos-Lerate, M.} and {Re Fiorentin, P.} and {Regibo, S.} and {Reylé, C.} and {Ripepi, V.} and {Riva, A.} and {Rixon, G.} and {Robichon, N.} and {Robin, C.} and {Roelens, M.} and {Rohrbasser, L.} and {Romero-Gómez, M.} and {Rowell, N.} and {Royer, F.} and {Rybicki, K. A.} and {Sadowski, G.} and {Sagristà Sellés, A.} and {Sahlmann, J.} and {Salgado, J.} and {Salguero, E.} and {Samaras, N.} and {Sanchez Gimenez, V.} and {Sanna, N.} and {Santoveña, R.} and {Sarasso, M.} and {Schultheis, M.} and {Sciacca, E.} and {Segol, M.} and {Segovia, J. C.} and {Ségransan, D.} and {Semeux, D.} and {Shahaf, S.} and {Siddiqui, H. I.} and {Siebert, A.} and {Siltala, L.} and {Slezak, E.} and {Smart, R. L.} and {Solano, E.} and {Solitro, F.} and {Souami, D.} and {Souchay, J.} and {Spagna, A.} and {Spoto, F.} and {Steele, I. A.} and {Steidelmüller, H.} and {Stephenson, C. A.} and {Süveges, M.} and {Szabados, L.} and {Szegedi-Elek, E.} and {Taris, F.} and {Tauran, G.} and {Taylor, M. B.} and {Teixeira, R.} and {Thuillot, W.} and {Tonello, N.} and {Torra, F.} and {Torra, J.} and {Turon, C.} and {Unger, N.} and {Vaillant, M.} and {van Dillen, E.} and {Vanel, O.} and {Vecchiato, A.} and {Viala, Y.} and {Vicente, D.} and {Voutsinas, S.} and {Weiler, M.} and {Wevers, T.} and {Wyrzykowski, Ł.} and {Yoldas, A.} and {Yvard, P.} and {Zhao, H.} and {Zorec, J.} and {Zucker, S.} and {Zurbach, C.} and {Zwitter, T.}},
	title = {Gaia Early Data Release 3 - Summary of the contents and survey properties},
	DOI= "10.1051/0004-6361/202039657",
	url= "https://doi.org/10.1051/0004-6361/202039657",
	journal = {A\&A},
	year = 2021,
	volume = 649,
	pages = "A1",
}

@article{Malo_2013,
doi = {10.1088/0004-637X/762/2/88},
url = {https://dx.doi.org/10.1088/0004-637X/762/2/88},
year = {2012},
month = {dec},
publisher = {The American Astronomical Society},
volume = {762},
number = {2},
pages = {88},
author = {Lison Malo and René Doyon and David Lafrenière and Étienne Artigau and Jonathan Gagné and Frédérique Baron and Adric Riedel},
title = {BAYESIAN ANALYSIS TO IDENTIFY NEW STAR CANDIDATES IN NEARBY YOUNG STELLAR KINEMATIC GROUPS*},
journal = {\apj}}

@article{Miles_2023,
doi = {10.3847/2041-8213/acb04a},
url = {https://dx.doi.org/10.3847/2041-8213/acb04a},
year = {2023},
month = {mar},
publisher = {The American Astronomical Society},
volume = {946},
number = {1},
pages = {L6},
author = {Brittany E. Miles and Beth A. Biller and Polychronis Patapis and Kadin Worthen and Emily Rickman and Kielan K. W. Hoch and Andrew Skemer and Marshall D. Perrin and Niall Whiteford and Christine H. Chen and B. Sargent and Sagnick Mukherjee and Caroline V. Morley and Sarah E. Moran and Mickael Bonnefoy and Simon Petrus and Aarynn L. Carter and Elodie Choquet and Sasha Hinkley and Kimberly Ward-Duong and Jarron M. Leisenring and Maxwell A. Millar-Blanchaer and Laurent Pueyo and Shrishmoy Ray and Steph Sallum and Karl R. Stapelfeldt and Jordan M. Stone and Jason J. Wang and Olivier Absil and William O. Balmer and Anthony Boccaletti and Mariangela Bonavita and Mark Booth and Brendan P. Bowler and Gael Chauvin and Valentin Christiaens and Thayne Currie and Camilla Danielski and Jonathan J. Fortney and Julien H. Girard and Carol A. Grady and Alexandra Z. Greenbaum and Thomas Henning and Dean C. Hines and Markus Janson and Paul Kalas and Jens Kammerer and Grant M. Kennedy and Matthew A. Kenworthy and Pierre Kervella and Pierre-Olivier Lagage and Ben W. P. Lew and Michael C. Liu and Bruce Macintosh and Sebastian Marino and Mark S. Marley and Christian Marois and Elisabeth C. Matthews and Brenda C. Matthews and Dimitri Mawet and Michael W. McElwain and Stanimir Metchev and Michael R. Meyer and Paul Molliere and Eric Pantin and Andreas Quirrenbach and Isabel Rebollido and Bin B. Ren and Glenn Schneider and Malavika Vasist and Mark C. Wyatt and Yifan Zhou and Zackery W. Briesemeister and Marta L. Bryan and Per Calissendorff and Faustine Cantalloube and Gabriele Cugno and Matthew De Furio and Trent J. Dupuy and Samuel M. Factor and Jacqueline K. Faherty and Michael P. Fitzgerald and Kyle Franson and Eileen C. Gonzales and Callie E. Hood and Alex R. Howe and Adam L. Kraus and Masayuki Kuzuhara and Anne-Marie Lagrange and Kellen Lawson and Cecilia Lazzoni and Pengyu Liu and Jorge Llop-Sayson and James P. Lloyd and Raquel A. Martinez and Johan Mazoyer and Sascha P. Quanz and Jea Adams Redai and Matthias Samland and Joshua E. Schlieder and Motohide Tamura and Xianyu Tan and Taichi Uyama and Arthur Vigan and Johanna M. Vos and Kevin Wagner and Schuyler G. Wolff and Marie Ygouf and Xi Zhang and Keming Zhang and Zhoujian Zhang},
title = {The JWST Early-release Science Program for Direct Observations of Exoplanetary Systems II: A 1 to 20 μm Spectrum of the Planetary-mass Companion VHS 1256–1257 b},
journal = {\apjl}}

@article{Balmer_2024, doi = {10.3847/1538-3881/ad1689}, url = {https://dx.doi.org/10.3847/1538-3881/ad1689}, year = {2024}, month = {jan}, publisher = {The American Astronomical Society}, volume = {167}, number = {2}, pages = {64}, author = {William O. Balmer and L. Pueyo and S. Lacour and J. J. Wang and T. Stolker and J. Kammerer and N. Pourré and M. Nowak and E. Rickman and S. Blunt and A. Sivaramakrishnan and D. Sing and K. Wagner and G.-D. Marleau and A.-M. Lagrange and R. Abuter and A. Amorim and R. Asensio-Torres and J.-P. Berger and H. Beust and A. Boccaletti and A. Bohn and M. Bonnefoy and H. Bonnet and M. S. Bordoni and G. Bourdarot and W. Brandner and F. Cantalloube and P. Caselli and B. Charnay and G. Chauvin and A. Chavez and E. Choquet and V. Christiaens and Y. Clénet and V. Coudé du Foresto and A. Cridland and R. Davies and R. Dembet and A. Drescher and G. Duvert and A. Eckart and F. Eisenhauer and N. M. Förster Schreiber and P. Garcia and R. Garcia Lopez and E. Gendron and R. Genzel and S. Gillessen and J. H. Girard and S. Grant and X. Haubois and G. Heißel and Th. Henning and S. Hinkley and S. Hippler and M. Houllé and Z. Hubert and L. Jocou and M. Keppler and P. Kervella and L. Kreidberg and N. T. Kurtovic and V. Lapeyrère and J.-B. Le Bouquin and P. Léna and D. Lutz and A.-L. Maire and F. Mang and A. Mérand and P. Mollière and C. Mordasini and D. Mouillet and E. Nasedkin and T. Ott and G. P. P. L. Otten and C. Paladini and T. Paumard and K. Perraut and G. Perrin and O. Pfuhl and D. C. Ribeiro and L. Rodet and Z. Rustamkulov and J. Shangguan and T. Shimizu and C. Straubmeier and E. Sturm and L. J. Tacconi and A. Vigan and F. Vincent and K. Ward-Duong and F. Widmann and T. Winterhalder and J. Woillez and S. Yazici and the GRAVITY Collaboration}, title = {VLTI/GRAVITY Provides Evidence the Young, Substellar Companion HD 136164 Ab Formed Like a “Failed Star”}, journal = {\aj} }

@article{Torres_2008,
       author = {{Torres}, C.~A.~O. and {Quast}, G.~R. and {Melo}, C.~H.~F. and {Sterzik}, M.~F.},
        title = "{Young Nearby Loose Associations}",
    booktitle = {Handbook of Star Forming Regions, Volume II},
         year = 2008,
       editor = {{Reipurth}, B.},
       volume = {5},
        pages = {757},
          doi = {10.48550/arXiv.0808.3362},
       adsurl = {https://ui.adsabs.harvard.edu/abs/2008hsf2.book..757T}}

@article{Brock_2021,
doi = {10.3847/1538-4357/abfc46},
url = {https://dx.doi.org/10.3847/1538-4357/abfc46},
year = {2021},
month = {jun},
publisher = {The American Astronomical Society},
volume = {914},
number = {2},
pages = {124},
author = {Laci Brock and Travis Barman and Quinn M. Konopacky and Jordan M. Stone},
title = {Cloud Properties of Brown Dwarf Binaries across the L/T Transition},
journal = {\apj}}

@article{Wang_2023,
doi = {10.3847/1538-3881/ac9f19},
url = {https://dx.doi.org/10.3847/1538-3881/ac9f19},
year = {2022},
month = {dec},
publisher = {The American Astronomical Society},
volume = {165},
number = {1},
pages = {4},
author = {Ji Wang and Jason J. Wang and Jean-Baptiste Ruffio and Geoffrey A. Blake and Dimitri Mawet and Ashley Baker and Randall Bartos and Charlotte Z. Bond and Benjamin Calvin and Sylvain Cetre and Jacques-Robert Delorme and Greg Doppmann and Daniel Echeverri and Luke Finnerty and Michael P. Fitzgerald and Nemanja Jovanovic and Ronald Lopez and Emily C. Martin and Evan Morris and Jacklyn Pezzato and Sam Ragland and Garreth Ruane and Ben Sappey and Tobias Schofield and Andrew Skemer and Taylor Venenciano and J. Kent Wallace and Peter Wizinowich and Jerry W. Xuan and Marta L. Bryan and Arpita Roy and Nicole L. Wallack},
title = {Retrieving C and O Abundance of HR 8799 c by Combining High- and Low-resolution Data},
journal = {\aj}}

@article{Hoch_2023,
doi = {10.3847/1538-3881/ace442},
url = {https://dx.doi.org/10.3847/1538-3881/ace442},
year = {2023},
month = {aug},
publisher = {The American Astronomical Society},
volume = {166},
number = {3},
pages = {85},
author = {Kielan K. W. Hoch and Quinn M. Konopacky and Christopher A. Theissen and Jean-Baptiste Ruffio and Travis S. Barman and Emily L. Rickman and Marshall D. Perrin and Bruce Macintosh and Christian Marois},
title = {Assessing the C/O Ratio Formation Diagnostic: A Potential Trend with Companion Mass},
journal = {\aj}}

@article{Cushing_2006,
doi = {10.1086/505637},
url = {https://dx.doi.org/10.1086/505637},
year = {2006},
month = {sep},
publisher = {},
volume = {648},
number = {1},
pages = {614},
author = {Michael C. Cushing and Thomas L. Roellig and Mark S. Marley and D. Saumon and S. K. Leggett and J. Davy Kirkpatrick and John C. Wilson and G. C. Sloan and Amy K. Mainzer and Jeff E. Van Cleve and James R. Houck},
title = {A Spitzer Infrared Spectrograph Spectral Sequence of M, L, and T Dwarfs},
journal = {\apj}}

@article{astropy_2013,
	author = {{The Astropy Collaboration} and {Robitaille, Thomas P.} and {Tollerud, Erik J.} and {Greenfield, Perry} and {Droettboom, Michael} and {Bray, Erik} and {Aldcroft, Tom} and {Davis, Matt} and {Ginsburg, Adam} and {Price-Whelan, Adrian M.} and {Kerzendorf, Wolfgang E.} and {Conley, Alexander} and {Crighton, Neil} and {Barbary, Kyle} and {Muna, Demitri} and {Ferguson, Henry} and {Grollier, Frédéric} and {Parikh, Madhura M.} and {Nair, Prasanth H.} and {Günther, Hans M.} and {Deil, Christoph} and {Woillez, Julien} and {Conseil, Simon} and {Kramer, Roban} and {Turner, James E. H.} and {Singer, Leo} and {Fox, Ryan} and {Weaver, Benjamin A.} and {Zabalza, Victor} and {Edwards, Zachary I.} and {Azalee Bostroem, K.} and {Burke, D. J.} and {Casey, Andrew R.} and {Crawford, Steven M.} and {Dencheva, Nadia} and {Ely, Justin} and {Jenness, Tim} and {Labrie, Kathleen} and {Lim, Pey Lian} and {Pierfederici, Francesco} and {Pontzen, Andrew} and {Ptak, Andy} and {Refsdal, Brian} and {Servillat, Mathieu} and {Streicher, Ole}},
	title = {Astropy: A community Python package for astronomy},
	DOI= "10.1051/0004-6361/201322068",
	url= "https://doi.org/10.1051/0004-6361/201322068",
	journal = {A\&A},
	year = 2013,
	volume = 558,
	pages = "A33",
	month = "",
}

@article{Husser2013,
	author = {Husser, T.-O. and {Wende-von Berg, S.} and {Dreizler, S.} and {Homeier, D.} and {Reiners, A.} and {Barman, T.} and {Hauschildt, P. H.}},
	title = {A new extensive library of PHOENIX stellar atmospheres and
          synthetic spectra},
	DOI= "10.1051/0004-6361/201219058",
	url= "https://doi.org/10.1051/0004-6361/201219058",
	journal = {A\&A},
	year = 2013,
	volume = 553,
	pages = "A6",
	month = "",
}

@ARTICLE{Asplund_2009,
       author = {{Asplund}, Martin and {Grevesse}, Nicolas and {Sauval}, A. Jacques and {Scott}, Pat},
        title = "{The Chemical Composition of the Sun}",
      journal = {\araa},
         year = 2009,
        month = sep,
       volume = {47},
       number = {1},
        pages = {481-522},
          doi = {10.1146/annurev.astro.46.060407.145222}}

@article{astropy_2018,
doi = {10.3847/1538-3881/aabc4f},
url = {https://dx.doi.org/10.3847/1538-3881/aabc4f},
year = {2018},
month = {aug},
publisher = {The American Astronomical Society},
volume = {156},
number = {3},
pages = {123},
author = {{The Astropy Collaboration} and A. M. Price-Whelan and B. M. Sipőcz and H. M. Günther and P. L. Lim and S. M. Crawford and S. Conseil and D. L. Shupe and M. W. Craig and N. Dencheva and A. Ginsburg and J. T. VanderPlas and L. D. Bradley and D. Pérez-Suárez and M. de Val-Borro and (Primary Paper Contributors) and T. L. Aldcroft and K. L. Cruz and T. P. Robitaille and E. J. Tollerud and (Astropy Coordination Committee) and C. Ardelean and T. Babej and Y. P. Bach and M. Bachetti and A. V. Bakanov and S. P. Bamford and G. Barentsen and P. Barmby and A. Baumbach and K. L. Berry and F. Biscani and M. Boquien and K. A. Bostroem and L. G. Bouma and G. B. Brammer and E. M. Bray and H. Breytenbach and H. Buddelmeijer and D. J. Burke and G. Calderone and J. L. Cano Rodríguez and M. Cara and J. V. M. Cardoso and S. Cheedella and Y. Copin and L. Corrales and D. Crichton and D. D’Avella and C. Deil and É. Depagne and J. P. Dietrich and A. Donath and M. Droettboom and N. Earl and T. Erben and S. Fabbro and L. A. Ferreira and T. Finethy and R. T. Fox and L. H. Garrison and S. L. J. Gibbons and D. A. Goldstein and R. Gommers and J. P. Greco and P. Greenfield and A. M. Groener and F. Grollier and A. Hagen and P. Hirst and D. Homeier and A. J. Horton and G. Hosseinzadeh and L. Hu and J. S. Hunkeler and Ž. Ivezić and A. Jain and T. Jenness and G. Kanarek and S. Kendrew and N. S. Kern and W. E. Kerzendorf and A. Khvalko and J. King and D. Kirkby and A. M. Kulkarni and A. Kumar and A. Lee and D. Lenz and S. P. Littlefair and Z. Ma and D. M. Macleod and M. Mastropietro and C. McCully and S. Montagnac and B. M. Morris and M. Mueller and S. J. Mumford and D. Muna and N. A. Murphy and S. Nelson and G. H. Nguyen and J. P. Ninan and M. Nöthe and S. Ogaz and S. Oh and J. K. Parejko and N. Parley and S. Pascual and R. Patil and A. A. Patil and A. L. Plunkett and J. X. Prochaska and T. Rastogi and V. Reddy Janga and J. Sabater and P. Sakurikar and M. Seifert and L. E. Sherbert and H. Sherwood-Taylor and A. Y. Shih and J. Sick and M. T. Silbiger and S. Singanamalla and L. P. Singer and P. H. Sladen and K. A. Sooley and S. Sornarajah and O. Streicher and P. Teuben and S. W. Thomas and G. R. Tremblay and J. E. H. Turner and V. Terrón and M. H. van Kerkwijk and A. de la Vega and L. L. Watkins and B. A. Weaver and J. B. Whitmore and J. Woillez and V. Zabalza and (Astropy Contributors)},
title = {The Astropy Project: Building an Open-science Project and Status of the v2.0 Core Package*},
journal = {\aj}}

@article{astropy_2022,
doi = {10.3847/1538-4357/ac7c74},
url = {https://dx.doi.org/10.3847/1538-4357/ac7c74},
year = {2022},
month = {aug},
publisher = {The American Astronomical Society},
volume = {935},
number = {2},
pages = {167},
author = {{The Astropy Collaboration} and Adrian M. Price-Whelan and Pey Lian Lim and Nicholas Earl and Nathaniel Starkman and Larry Bradley and David L. Shupe and Aarya A. Patil and Lia Corrales and C. E. Brasseur and Maximilian Nöthe and Axel Donath and Erik Tollerud and Brett M. Morris and Adam Ginsburg and Eero Vaher and Benjamin A. Weaver and James Tocknell and William Jamieson and Marten H. van Kerkwijk and Thomas P. Robitaille and Bruce Merry and Matteo Bachetti and H. Moritz Günther and Paper Authors and Thomas L. Aldcroft and Jaime A. Alvarado-Montes and Anne M. Archibald and Attila Bódi and Shreyas Bapat and Geert Barentsen and Juanjo Bazán and Manish Biswas and Médéric Boquien and D. J. Burke and Daria Cara and Mihai Cara and Kyle E Conroy and Simon Conseil and Matthew W. Craig and Robert M. Cross and Kelle L. Cruz and Francesco D’Eugenio and Nadia Dencheva and Hadrien A. R. Devillepoix and Jörg P. Dietrich and Arthur Davis Eigenbrot and Thomas Erben and Leonardo Ferreira and Daniel Foreman-Mackey and Ryan Fox and Nabil Freij and Suyog Garg and Robel Geda and Lauren Glattly and Yash Gondhalekar and Karl D. Gordon and David Grant and Perry Greenfield and Austen M. Groener and Steve Guest and Sebastian Gurovich and Rasmus Handberg and Akeem Hart and Zac Hatfield-Dodds and Derek Homeier and Griffin Hosseinzadeh and Tim Jenness and Craig K. Jones and Prajwel Joseph and J. Bryce Kalmbach and Emir Karamehmetoglu and Mikołaj Kałuszyński and Michael S. P. Kelley and Nicholas Kern and Wolfgang E. Kerzendorf and Eric W. Koch and Shankar Kulumani and Antony Lee and Chun Ly and Zhiyuan Ma and Conor MacBride and Jakob M. Maljaars and Demitri Muna and N. A. Murphy and Henrik Norman and Richard O’Steen and Kyle A. Oman and Camilla Pacifici and Sergio Pascual and J. Pascual-Granado and Rohit R. Patil and Gabriel I Perren and Timothy E. Pickering and Tanuj Rastogi and Benjamin R. Roulston and Daniel F Ryan and Eli S. Rykoff and Jose Sabater and Parikshit Sakurikar and Jesús Salgado and Aniket Sanghi and Nicholas Saunders and Volodymyr Savchenko and Ludwig Schwardt and Michael Seifert-Eckert and Albert Y. Shih and Anany Shrey Jain and Gyanendra Shukla and Jonathan Sick and Chris Simpson and Sudheesh Singanamalla and Leo P. Singer and Jaladh Singhal and Manodeep Sinha and Brigitta M. Sipőcz and Lee R. Spitler and David Stansby and Ole Streicher and Jani Šumak and John D. Swinbank and Dan S. Taranu and Nikita Tewary and Grant R. Tremblay and Miguel de Val-Borro and Samuel J. Van Kooten and Zlatan Vasović and Shresth Verma and José Vinícius de Miranda Cardoso and Peter K. G. Williams and Tom J. Wilson and Benjamin Winkel and W. M. Wood-Vasey and Rui Xue and Peter Yoachim and Chen Zhang and Andrea Zonca and Astropy Project Contributors},
title = {The Astropy Project: Sustaining and Growing a Community-oriented Open-source Project and the Latest Major Release (v5.0) of the Core Package*},
journal = {\apj}}

@article{Barenfeld_2013,
doi = {10.1088/0004-637X/766/1/6},
url = {https://dx.doi.org/10.1088/0004-637X/766/1/6},
year = {2013},
month = {feb},
publisher = {The American Astronomical Society},
volume = {766},
number = {1},
pages = {6},
author = {Scott A. Barenfeld and Eric J. Bubar and Eric E. Mamajek and Patrick A. Young},
title = {A KINE-CHEMICAL INVESTIGATION OF THE AB DOR MOVING GROUP “STREAM”},
journal = {\apj}}

@misc{ian_czekala_2018_2221006,
  author       = {Ian Czekala and
                  Michael Gully-Santiago and
                  Kevin Gullikson and
                  Sean Andrews and
                  Jason Neal and
                  Miles Lucas and
                  Kevin Hardegree-Ullman and
                  Meredith Rawls and
                  Edward Betts},
  title        = {{iancze/Starfish: ca. Czekala et al. 2015 release 
                   w/ Zenodo}},
  month        = dec,
  year         = 2018,
  publisher    = {Zenodo},
  version      = {v0.2.2},
  doi          = {10.5281/zenodo.2221006},
  url          = {https://doi.org/10.5281/zenodo.2221006}
}

@article{Parviainen2015,
  author = {Parviainen, Hannu and Aigrain, Suzanne},
  doi = {10.1093/mnras/stv1857},
  journal = {MNRAS},
  month = nov,
  number = {4},
  pages = {3821--3826},
  title = {{ldtk: Limb Darkening Toolkit}},
  url = {http://mnras.oxfordjournals.org/lookup/doi/10.1093/mnras/stv1857},
  volume = {453},
  year = {2015}
}
\bibliographystyle{aasjournalv7}

\end{document}